\documentclass{aa}  
\usepackage{graphicx}
\usepackage{multirow}
\usepackage[draft,inline,nomargin,index]{fixme}
\usepackage{subcaption}
\usepackage[normalem]{ulem}
\usepackage[colorlinks=true,linkcolor=blue,citecolor=blue,urlcolor=blue,breaklinks=true]{hyperref}

\usepackage{txfonts}

\fxsetup{theme=color,mode=multiuser}

\definecolor{necolor}{RGB}{197,66,245}

\definecolor{denis_color}{RGB}{255,0,0}

\definecolor{armando_color}{RGB}{0,128,0}

\definecolor{philippe_color}{RGB}{252,127,3}

\definecolor{kat_color}{RGB}{3,244,252}

\definecolor{olc_color}{RGB}{3,243,252}

\newcommand{\teff}{\ensuremath{T_{\mathrm{eff}}}}
\newcommand{\logg}{\log g}
\newcommand{\msol}{M\ensuremath{_\odot}}

\begin{document} 

   \title{Interferometric Survey of Stellar Parameters: Limb darkening study with the "Pipeline for Interferometric Measurements of Stars" (PIMS) }
    \titlerunning{Spectroscopy, interferometry and atmosphere models}
    \authorrunning{Ebrahimkutty et al.}

   \author{ N.~Ebrahimkutty\inst{1} \and D.~Mourard\inst{1} \and P.~Berio\inst{1} \and O.~Creevey\inst{1} \and A.~Domiciano~de~Souza\inst{1} \and K.~Lee\inst{2} \and T.~Morel\inst{3} \and B.~Plez \inst{4} \and D.~Salabert\inst{1} \and N.~Anugu\inst{5} \and R.~V.~Iba\~{n}ez~Bustos\inst{1} \and J.~Jon\'{a}k\inst{1} \and H.~Nowacki\inst{1} \and M.~Vrard\inst{1} \and J.~Dejonghe\inst{1} \and F.~Morand\inst{1} \and N.~Nardetto\inst{1} \and A.~Meilland\inst{1} \and K.~Perraut\inst{6} \and M.~Wittkowski\inst{7} \and J.~Monnier\inst{8} \and S.~Kraus\inst{9} \and M.~Gutierrez\inst{8} \and N.~Ibrahim\inst{8} }

   \institute{Universit\'e C\^ote d'Azur, Observatoire de la C\^ote d'Azur, CNRS, Laboratoire Lagrange, France\\
              \email{nayeem.ebrahimkutty@oca.eu}     
         \and
            Max Plank Institute for Astronomy, 69117 Heidelberg, Germany
         \and
                 Space Sciences, Technologies and Astrophysics Research (STAR) Institute, Université de Liège, Quartier Agora, Allée du 6 Août 19c, Bât. B5C, 4000 Liège, Belgium
        \and
            LUPM, Univ Montpellier, CNRS, Montpellier, France
        \and
            The CHARA Array of Georgia State University, Mount Wilson Observatory, Mount Wilson, CA 91203, USA 
        \and
            Univ. Grenoble Alpes, CNRS, IPAG, 38000 Grenoble, France
        \and
            European Southern Observatory, Karl-Schwarzschild-Str. 2, 85748 Garching bei München, Germany  
        \and
            Astronomy Department, University of Michigan, Ann Arbor, MI 48109, USA
        \and 
            Astrophysics Group, Department of Physics \& Astronomy, University of Exeter, Stocker Road, Exeter, EX4 4QL, UK
        }

   \date{Received ....; accepted ...}

  \abstract
   {The accurate and unbiased determination of stellar parameters is of importance in many astrophysical domains.}
   {We present an improved method to estimate stellar parameters, based on interferometric observations and stellar atmosphere models, along with spectroscopic and photometric data.}
   {The fundamental stellar parameters are estimated by fitting a model of intensity profiles to interferometric data over a large spectral window and combining them with spectroscopic and photometric observations as an additional constraint on the fitting. We further present results obtained with this method using observations from the Center for High Angular Resolution Astronomy (CHARA) and archival data from the Very Large Telescope Interferometer (VLTI). Based on discrete grids of one-dimensional (1D) stellar atmosphere models, we developed an algorithm that trains artificial neural networks (ANNs), capable of estimating the spectrum and intensity profile of a star over the three photometric bands $R$, $H$, and $K$. The ANN cover effective temperatures ($T_\mathrm{eff}$) ranging from 4500–7000 K for dwarf stars and 2500–8000 K for giant stars, with surface gravities ($\log g$) ranging from 3.0–5.0 dex and $-0.5$–3.5 dex, respectively, and 12 viewing angles. A $\chi^2$ minimization algorithm based on trained functions permits us to simultaneously fit the observational spectrum and interferometric complex visibilities. As a result, consistent and precise stellar parameters, such as $\teff$, $\logg$, and angular diameter ($\theta$), can be estimated. }
   {We fitted $\iota$ Psc (F7V) and $\delta$ Ari (G9.5IIIb) observed with SPICA, MIRC-X, and MYSTIC at the CHARA Array and on archival data of $\alpha$ Cen A and B (G2V and K1V) from VLTI/PIONIER with PIMS. For angular diameter estimations, fitting with a one-dimensional stellar atmosphere model, we reached a precision of 0.5\%. For $\teff$ and radius, we obtain a precision of approximately 1\%,  and 5\% precision for $\logg$. Thanks to evolutionary models, we further show how these constraints can improve the mass and age determination, attaining a precision of approximately 2\% for mass and 5-10\% for age.}  
   {This paper demonstrates the first results from the model fitting approach using interferometric data from different instruments combined with spectroscopic observations.}


   \maketitle
%

\section{Introduction}
\label{sec:1}

Optical interferometry provides a direct means of measuring stellar angular diameters by fitting analytical descriptions of the intensity distribution across the stellar disk, a technique explored in numerous studies \citep{Brown1974, Diaz-Cordoves1992, Claret2000, Davis2000, Claret2022}. Once the angular diameter is known, additional stellar parameters, such as stellar radii and effective temperature, can be inferred using standard relations. Additionally, \citet{Wittkoski2004}, \citet{Aufdenberg2005}, and \citet{Wittkowski2006} have demonstrated the use of fitting interferometric and spectroscopic data with stellar atmosphere models to constrain the models and estimate stellar parameters. Furthermore, combining these with stellar evolution models permits strong constraints on the masses and ages of the star.

A key aspect of these interferometric analyses is the description of the variation in stellar intensity from center to limb, called limb-darkening (LD). It is commonly expressed as a function of $\mu = \cos \alpha$, where $\alpha$ is the angle between the line of sight and the normal to the local surface. Intensity profiles are typically approximated by analytical LD laws with a small number of coefficients \citep{Claret2000, Espinoza2016, Morello2020}. Accurate LD prescriptions are required not only for interferometric diameter measurements but also for the interpretation of exoplanet transit light curves \citep{Maxted2018}. Therefore, stellar atmosphere models have been widely used to predict LD coefficients and to test LD laws against observations \citep{Claret2000, Kervella2017}. In particular, \citet{Kervella2017} compared various LD laws from different stellar atmosphere models with observations and found that the square root law or a four-parameter law provided the best fit, although they were not significantly more accurate than the single-parameter power law.

The Interferometric Survey of Stellar Parameters (ISSP), conducted with the CHARA/SPICA instrument \citep{Mourard2026} along with the MIRC-X \citep{Anugu2020} and MYSTIC \citep{Setterholm2023} instruments, is currently ongoing and will provide a homogeneous set of limb-darkened angular diameter ($\theta_{\rm{LD}}$) measurements for several hundred stars. With a resolution down to $0.2$ mas, CHARA/SPICA enables direct LD constraints for a wide range of stars across the Hertzsprung–Russell (HR) diagram. 

This survey is particularly relevant for the PLAnetary Transits and Oscillations of stars (PLATO) mission \citep{Rauer2025}, which requires stellar radii to be determined with a precision of a few per cent for its primary targets. Within PLATO, two different pipelines have been developed to determine various stellar properties: the Module Stellar Science pipelines (MSteSci's) and the Module Stellar Analysis Pipelines (MSAPs). One of the Stellar Science pipelines, MSteSci1, will operate prior to satellite operations and will process preparatory data mainly obtained from ground-based observations. A part of MSteSci1 employs the interferometric method described in \citet{Ebrahimkutty2024} (hereafter referred to as Paper $1$). MSteSci1 will provide initial estimates of stellar parameters such as angular diameter, effective temperature, and radius for stars included in the PLATO Input Catalog \citep[PIC;][]{Montalto2026}.

In Paper $1$, we introduced a method that combines interferometric, spectroscopic, and photometric data with Model Atmospheres with a Radiative and Convective Scheme (MARCS) stellar atmosphere models \citep{Gustafsson2008} and demonstrated that the square-root LD law provides the best approximation to these models. In this work, we also provided details of the basic architecture of the analysis pipeline. This approach enabled the simultaneous estimation of the angular diameter ($\theta$), effective temperature ($\teff$), and surface gravity ($\logg$). The stellar radius is derived from $\theta$ using the parallax. However, Paper $1$ was limited to low spectral resolution CHARA/SPICA data with fixed spectral sampling, and the artificial neural network (ANN) used for the analysis was trained only on plane-parallel MARCS models, restricting its scope to FGK dwarfs and subgiants. 
In the work presented here, we expand the scope of the fitting process to include giant stars and provide a framework that incorporates multi-band observations in the $R$, $H$, and $K$ bands. This will allow for fitting either combined observations or individual measurements.

In Sect.~\ref{sec:2}, we first outline the improvements made to the interferometric pipeline, which we refer to as the Pipeline for Interferometric Measurements of Stars (\texttt{PIMS}). We also discuss the additional work done on the propagation of uncertainties in the interferometric data. Sect.~\ref{sec:3} presents the actual data set used to demonstrate the potential of the \texttt{PIMS}. The results and discussions are presented in Sect.~\ref{sec:4}. The \texttt{PIMS} is integrated into the MSteSci1 pipeline; this implementation is explained in Sect. \ref{sec:5}. In section~\ref{sec:6}, as a simple application, we demonstrate how the $\teff$, $\log g$ and radius estimated from the interferometric data can be combined with stellar evolution models for an accurate determination of age and mass. Sect.~\ref{sec:7} gives our conclusions.

\section{Improved neural network and pipeline}
\label{sec:2}

The current study extends the methodology in several directions. The framework is generalized to accommodate interferometric data from multiple instruments and spectral configurations. These include observations in $R$, $H$, and $K$ bands. The neural network training is expanded to incorporate MARCS models in spherical geometry, allowing the analysis of giant stars. These improvements increase both the efficiency and applicability of the method and establish a versatile tool for the coming era of large-scale interferometric surveys.

\subsection{Stellar atmosphere models}
\label{sec:2.1}

The MARCS models \citep{Gustafsson2008} are based on one-dimensional hydrostatic equilibrium and local thermodynamic equilibrium (LTE). We use plane-parallel geometries for dwarfs and spherical geometries for giants to compute the intensity profiles in the $R$, $H$, and $K$ bands. All models were computed assuming solar metallicity, as described in Paper~1. The intensity profiles are derived from the MARCS models using the non-local thermodynamic equilibrium (NLTE) version of the spectrum synthesis code Turbospectrum \citep[TS;][]{plez2012,gerber2023}. Table \ref{tab:table1} gives the range of parameters used in the model grid for each model geometry.

\begin{table}[ht!]
    \caption{Stellar atmosphere models grid}
    \small
    \centering
    \begin{tabular}{c c c}
    \hline
    Parameter & Plane-parallel & Spherical \\
    \hline
    $\teff$ (K) & 4500 -- 7000 & 2500 -- 8000 \\
    $\teff$ (K) step size & 100 -- 200 & 100 -- 250 \\
    $\log g$ (dex) & 3.0 -- 5.0 & -0.5 -- 3.5 \\
    $\log g$ (dex) step size & 0.5 & 0.5 \\
     R band range (nm) & 400 -- 900 & 400 -- 900 \\
     H band range (nm) & 1200 -- 1850 & 1200 -- 1850 \\
     K band range (nm) & 1950 -- 2400 & 1950 -- 2400 \\
     Initial resolution & 300000 & 300000 \\
    Training resolution & 500 & 500 \\
    \hline
    \end{tabular}

    \label{tab:table1}
\end{table}

Our pipeline integrates the spectroscopic and photometric modules adapted from the Stellar Abundances and Atmospheric Parameters Pipeline \citep[SAPP;][]{gent2022}.

To define the LD variation, the description of the intensity profiles is based on a Radau sampling with $12$ points at all wavelengths. In Paper 1, we showed that, for interferometric observations in the $R$ band, a 12-point Radau sampling is adequate for angular diameters smaller than 3 mas. We did a similar calculation for the $H$ and $K$ bands, and this sampling can be applied to angular diameters smaller than 5.5 mas and 8.75 mas in the $H$ and $K$ bands, respectively.

\subsection{Machine learning for intensity profiles}
\label{sec:2.2}

Since most of the interferometric observations are performed at low spectral resolution (50-140), the training of the ANN and the computation of intensity profiles benefit from a reduction in wavelength sampling with respect to what MARCS provides. To this end, the intensity profiles are binned to 500 spectral channels in each band, which permits accelerating the training while retaining sufficient spectral information. Observations from MIRC-X and MYSTIC show that the number of wavelength channels may vary, following the actual spectral alignment; the adopted binning strategy ensures that the ANN remains adaptable to these variations. The intensity profiles generated from the MARCS models were used to train the ANN separately for each geometry (plane-parallel and spherical) and each wavelength band, following the methodology outlined in Paper $1$.

From the trained ANN, the intensity profile of any given star can be predicted within the model grid limits (see Table~\ref{tab:table1}). These ANN-generated profiles are then interpolated and binned to match the wavelength sampling of the observations, as extracted from the corresponding FITS files. The model visibilities are computed by performing a Hankel transform (HT) of the one-dimensional intensity profile, as explained in Paper $1$.  The square visibility thus obtained provides the basis for fitting interferometric observations.

\subsection{Model fitting}
\label{sec:2.3}

As in Paper $1$, to estimate the stellar parameters, we combine interferometry with spectroscopy and photometry. Using the results from the spectroscopic and photometric modules, we can limit the number of models used in the interferometric module, while maintaining a reliable estimation of $\teff$ and $\logg$ and reaching better accuracy. 

The interferometric module needs starting values for each free parameter. We use $\teff$ and $\logg$ obtained from the spectroscopic and photometric modules, respectively. The initial value for $\theta$ is arbitrarily set to $1$ mas.

The initial fit to the data is performed using a chi-square ($\chi^2$) minimization and \texttt{scipy.optimize} routines in Python, treating $\theta$, $\teff$, and $\logg$ as free parameters. From this first estimation, the fitting of the model is refined with a Markov Chain Monte Carlo (MCMC) sampler using emcee \citep{forman2013PASP..125..306F} permitting to explore the posterior probability distributions of the parameters. Estimated values (e.g. the median) and uncertainties of the parameters are extracted from these posterior distribution functions (PDFs).

One of the inputs in the photometric module is the parallax (\( \pi_p \)) obtained from Gaia DR3 \citep{gaiadr3} or from \citet{kervella2016}. The photometric module calculates the distance as $1/\pi_p$. For the parallax from Gaia DR3, we have corrected for the parallax bias as described in \citet{Lindegren2021}. This distance is then used in the interferometric module to estimate the stellar radius using the relationship between ($ \theta $) and distance in parsecs. The uncertainty in the radius is determined using the standard error propagation method.

In the updated version of the pipeline, the $\chi^2$ statistic is defined as the sum of contributions from the spectroscopic module, the photometric module, and the square visibilities in the $R$, $H$, and $K$ bands, with equal weight. 

\subsection{Estimation of uncertainties}
\label{sec:2.4}

In paper $1$, we used simulated square visibilities created from MARCS models and mimicked observations from CHARA/SPICA using Aspro2 \footnote{Available at http://www.jmmc.fr/aspro} \citep{Bourg2016}. To reproduce the noise in a typical observed SPICA data set, we set a signal-to-noise ratio of 20 on the square visibility. Based on this estimation of the statistical noise, we build a Gaussian distribution to estimate a random term added to the computed visibility as a bias. Despite this, the uncertainties in the final fit from the MCMC exhibited very small values, probably not realistic.

\citet{Perrin2003} demonstrated through examples from the CHARA/FLUOR\footnote{CHARA/FLUOR is a beam combiner operating in the near-infrared K band \citep{Coule2003}. } instrument that there is a significant bias arising from the calibration data, highlighting the need to consider the correlation between them. The author also indicated that the error in square visibility cannot be treated directly, as there is a correlation between the calibrator and the transfer functions, since the same calibrator is used for both. Then, an analytical method to propagate the correlated uncertainties of square visibility was proposed.

Building on this, \citet{Lachaume2019} showed that more reliable uncertainties in interferometric observations can be achieved by applying a correlated statistical error and accounting for systematic errors using a bootstrap method. These correlations include the non-Gaussian statistical error from the raw visibility, from the determination of the transfer function, and from the calibrated visibilities.

For interferometric observations, it is important to account for the correlation between different baselines of the observations for both calibrators and targets throughout the night. \citet{Mourard2026}, have implemented a similar method to calculate the correlation matrix for each of the targets observed within the ISSP program and save it in the header of the oifits files.

With the correlation matrix, one can calculate the covariance using the following equation:

\begin{equation}
    \mathrm{C}_{ij} = \rho_{ij} \sigma_{i}\sigma_{j},
\end{equation}
where $\sigma$ is the uncertainty in the square visibility and $\rho$ is the correlation matrix. The indices $i$ and $j$ label individual interferometric measurements, such that $\mathrm{C}_{ij}$ represents the covariance between the errors of the $i$-th and $j$-th data points. The covariance matrix is regularized by adding a small diagonal term ($10^{-12}$) to ensure numerical stability during inversion. Taking into account the covariance matrix, C, in the fitting process of the interferometric module permits us to reach a better and more reliable $\chi^2$ statistic.

\begin{equation}
    \chi^2 = \mathrm{r}^\mathrm{T}\,\mathrm{C}^{-1}\,\mathrm{r},
\end{equation}
where $r$ is the residual between the observed and the modeled square visibilities. To compute the $\mathrm{C}^{-1}$, we use Cholesky decomposition using the functions from the \texttt{scipy.linalg} \citep{scipy} library.

Even after the use of the covariance matrix, one can obtain a reduced chi-square ($\chi^2_\textrm{red}$) value greater than 1. This indicates that the measurement uncertainties have been underestimated. In such cases, we can correct it by rescaling the final uncertainties,  multiplying them by the square root of the $\chi^2_{\mathrm{red}}$ \citep{Teunissen2008}.

\section{Implementation of the interferometric module in MSteSci1}
\label{sec:5}

\begin{figure*}
    \centering
    \includegraphics[width =\textwidth]{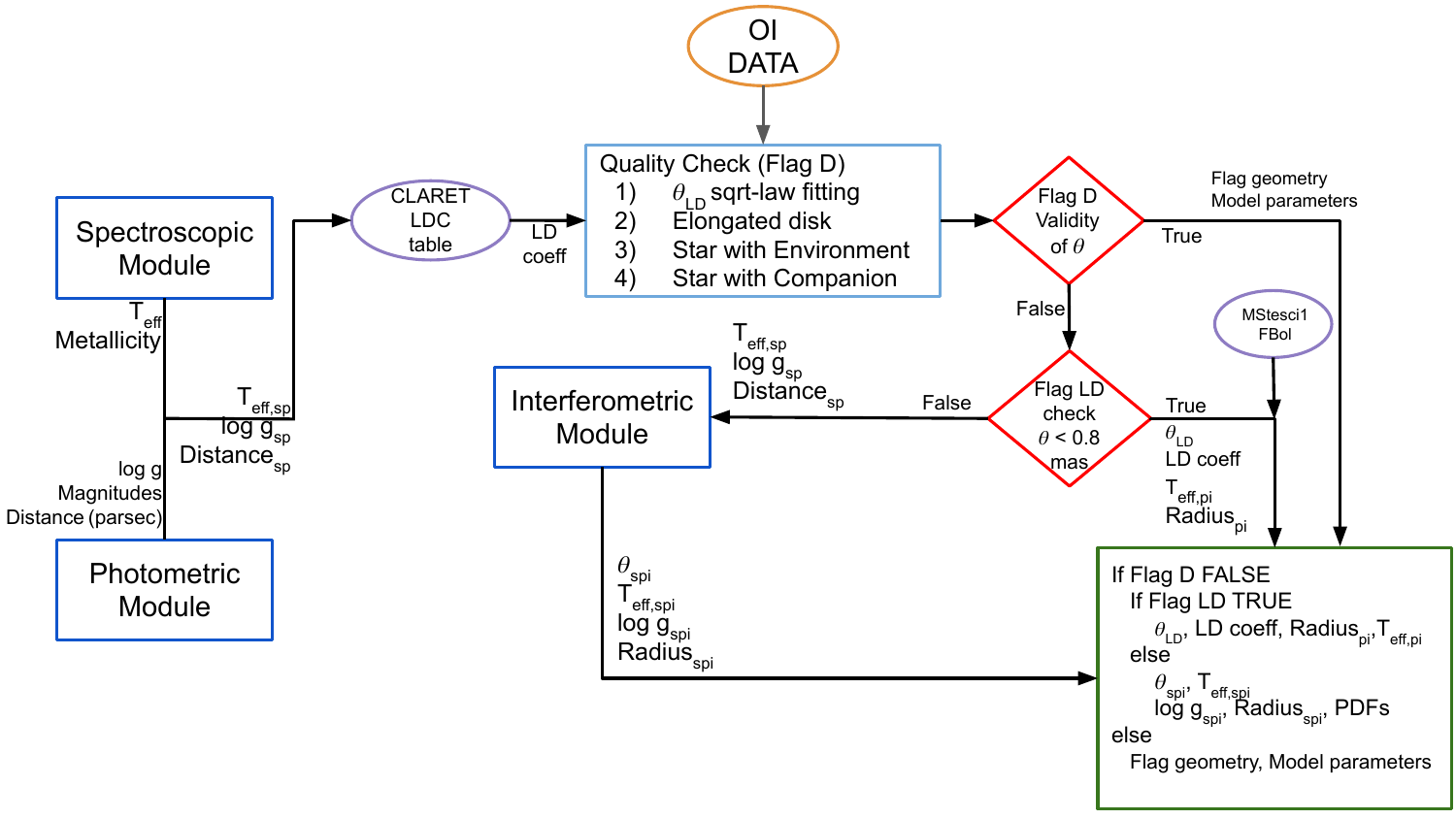}
        
    \caption{Flow chart representing the implementation of the interferometric module within the PLATO MSteSci1 pipeline. In the diagram, subscripts \textit{sp} refer to parameters from the spectroscopic module and the photometric module taken as input in the interferometric module, \textit{pi} to output parameters that are derived using the analytical method, \textit{spi} to output parameters obtained from the combined method, and \textit{LD} to limb-darkening.}
    \label{fig:fig1}
\end{figure*}

As mentioned in the introduction, the interferometric module presented here is integrated into the PLATO MSteSci1 pipeline. Figure \ref{fig:fig1} illustrates how the interferometric module is incorporated into MSteSci1. We use the $\teff$ and $\logg$ (marked as $T_\textrm{eff, sp}$ and $\textrm{log g}_{sp}$ in the flowchart) from the spectroscopic and photometric modules as inputs to the interferometric module, as detailed above. The blue box labeled "Interferometric Module" represents the algorithm explained in this paper and in Paper $1$.

The first step is to perform a quality check (represented in a light blue box in Fig. \ref{fig:fig1}) to determine if the target star is correctly described by a limb-darkened disk. To reduce computational cost, the coefficients of the square-root law used in this step are queried, for each spectral band, from the Claret LD catalogue \citep{Claret2011} using the $T_\mathrm{eff, sp}$ and $\textrm{log g}_{sp}$ values. To ensure consistency in the queried values, we restrict the coefficients to those obtained using the Least Squares Method (LSM) from ATLAS models.

Using these coefficients, we compute the analytical square visibility and the closure phase using the square-root limb-darkening law. We then fit the analytical square visibility and closure phase with the observations to see if the star has any deviation from the LD disc. Next, we apply least-squares minimisation on a Gaussian function and an elliptical function to check for the presence of a circumstellar environment and elongation in the star. Finally, we look if it is binary.

In all cases, the minimisation is performed using the $\chi^2_{\mathrm{red}}$. The resulting $\chi^2_{\mathrm{red}}$ values are compared across models to identify the most appropriate interpretation of the data, and the corresponding stellar classification is used to assign the flag ‘FLAG~D’. 'FLAG~D' is set to TRUE if the best model is not the LD disk. If 'FLAG D' is FALSE, which means that the data are compatible with an LD disk, we verify if the expected angular diameter is appropriate for constraining both the diameter and the limb darkening ($\theta > 0.8$ mas for CHARA/SPICA data). If this condition is not met, we flag 'FLAG LD' as TRUE and proceed to a square-root LD fit to obtain the angular diameter ($\theta_\textrm{LD}$). In addition, we use the bolometric flux (FBol), calculated in MSteSci1, to derive $\teff$ ($T_{\mathrm{eff,\,pi}}$) and using distance, we calculate the radius of the star.

If both flags, FLAG D and FLAG LD, are FALSE, we proceed to the interferometric module using the $T_\textrm{eff, sp}$ and $\textrm{log g}_{sp}$, along with the distance to the target in parsecs from MSteSci1, as inputs. The interferometric module then provides the values of $\theta$, $\teff$, $\logg$, and radius as the output of the combined method, which are denoted as $\theta_\mathrm{spi}$, $T_\textrm{eff, spi}$, $\textrm{log g}_\textrm{spi}$, and $\textrm{Radius}_\textrm{spi}$ in the flowchart shown in Figure \ref{fig:fig1}.

\section{Observations and data reduction}
\label{sec:3}

We observed HD 222368 ($\iota$ Psc, F7V) and HD 19787 ($\delta$ Ari, G9.5IIIb) using the SPICA \citep{Mourard2026}, MIRC-X \citep{Anugu2020}, and MYSTIC \citep{Setterholm2023} instruments at the CHARA Array. The instruments operate in the $R$ band (600–900 nm), $H$ band (1450-1700 nm), and $K$ band (1950–2400 nm), respectively. These observations were carried out as part of the S05 programme of the ongoing ISSP survey \citep{Mourard2026}, which is dedicated to estimating limb-darkened stellar parameters. The calibration of these data from all three instruments is done using the calibration tools developed for the SPICA observations \citep{Mourard2026}. Table \ref{tab:table2} lists the details of the interferometric observations used in this study along with the calibrators observed. 

\begin{table*}[]
    \caption{Interferometric data}
    \small
    \centering
    \begin{tabular}{c c c c c c c}
    \hline
    \shortstack[c]{Star} & \shortstack[c]{Stellar \\ Type} & \shortstack[c]{Date of \\ Observations} & \shortstack[c]{Instrument} & \shortstack[c]{No. of \\ Observations} & \shortstack[c]{Wavelength \\ Band} & \shortstack[c]{Calibrator} \\    
    \hline
    \multirow{3}{*}{HD 222368 ($\iota$ Psc)} & 
    \multirow{3}{*}{F7V} & 
    \multirow{3}{*}{15-09-2025} & 
    CHARA/SPICA  & 1 & R & \multirow{3}{*}{HD 220825, HD 7804} \\
    & & & CHARA/MIRCX & 1 & H & \\
    & & & CHARA/MYSTIC & 1 & K & \\
    \hline
    \multirow{3}{*}{HD 19787 ($\delta$ Ari)} & 
    \multirow{3}{*}{G9.5IIIb} & 
     & 
    CHARA/SPICA  & 1 & R & \multirow{3}{*}{HD 20150, HD 27045} \\
    & & 04, 05-10-2025 & CHARA/MIRCX & 1 & H & \\
    & &  & CHARA/MYSTIC & 1 & K & \\
    \hline
    HD128620 ($\alpha$ Cen A) & G2V & 23, 27, 29, 30-05-2016 & VLTI/PIONIER & 5 & H & HD 127753, HD 133869 \\
    \hline
    HD 128621 ($\alpha$ Cen B) & K1V & 23, 27, 29, 30-05-2016 & VLTI/PIONIER & 4 & H & HD 127753, HD 133869 \\
    \hline
         
    \end{tabular}
    \tablefoot{HD 222368 and HD 19787 are observed as part of the ISSP program, and the values of the correlation matrix are stored in their header. For HD 128620 and HD 128621, the calibrated data are obtained from the OiDB database.} 
    \label{tab:table2}
\end{table*}

All calibrators were selected using Night Scheduling Software \citep[NSS;][]{Mourard2026}, with well-determined angular diameters. The uniform disk angular diameter of the calibrators for each band is taken from \texttt{getstar\footnote{Available at https://www.jmmc.fr/getstar}}. They are observed in close temporal and angular proximity to the science targets to ensure accurate calibration. The resulting calibrated data are available through the OiDB\footnote{OiDB available at http://oidb.jmmc.fr} database \citep{Haubois2014SPIE.9146E..0OH}. The information regarding the calibrators used in each of the observations, for each band, along with the angular diameter of the calibrators used are provided in the headers of the data available in the OiDB. Before calibration, we analysed the square visibilities and closure phases of the calibrators to check for signatures of binarity or stellar activity, and no such signatures were detected.

For the SPICA observations of HD\,222368, we noticed a calibration issue, as all square visibilities were consistently higher than expected. After checking the calibrators, we interpret this as an effect of the magnitude difference between the target and the calibrators. To account for this effect, we used an analytical model fitting to adjust a background component and rescale the square visibilities prior to complete PIMS analysis.

To check the robustness of the \texttt{PIMS}, we decided, as an example to retrieve and analyse the calibrated interferometric data of $\alpha$ Cen A\&B from OiDB obtained with VLTI/PIONIER, as used in \citet{Kervella2017}. 

For the spectroscopic analysis, we retrieved spectra from the ESO Science Archive corresponding to observations with the Ultraviolet and Visual Echelle Spectrograph \citep[UVES;][]{Dekkar2000SPIE.4008..534D} and the High Accuracy Radial velocity Planet Searcher \citep[HARPS;][]{mayor2003Msngr.114...20M}. We performed the continuum normalisation and radial velocity corrections of these spectra using the open-source software \texttt{iSpec} \citep{Blanco-Cuaresma2019}.

Since the photometric module requires at least a few magnitude values for each target, we extracted available measurements from the Gaia DR3 catalogue \citep{gaiadr3} and queried Simbad \citep{simbad} for the other photometric values.

The reference parameters for the spectroscopic and photometric modules,  the $\teff$, $\logg$, and abundances, were adopted from the GSP-Spec in Gaia DR3 catalogue \citep{Creevey2023}, except for the $\alpha$ Cen stars, for which the values were taken from \cite{gent2022}. For $\pi_p$ were adopted from the Gaia DR3 catalogue, while for $\alpha$ Cen, they were taken from \citet{kervella2016}. Table \ref{tab:table3} lists the details of the spectroscopic data, magnitude values and the reference parameters used in this work.

\begin{table*}[]
    \caption{Reference, photometric, and spectroscopic data for the targets.}
    \small
    \centering
    \setlength{\tabcolsep}{5pt} 
    \renewcommand{\arraystretch}{1.2} 
    \begin{tabular}{lcccc}
    \hline
        & $\iota$ Psc & $\delta$ Ari & $\alpha$ Cen A & $\alpha$ Cen B \\
        & HD 222368 & HD 19787 & HD 128620 & HD 128621 \\
    \hline
    \multicolumn{5}{c}{\textbf{Reference Data}} \\
    \hline
    $T_{\mathrm{eff}}$ (K) & $6149 \pm 35$ & $4814 \pm 2$ & $5792 \pm 16$ & $5231 \pm 20$ \\
    $\log g$ (dex) & $3.930 \pm 0.230$ & $2.560 \pm 0.010$ & $4.310 \pm 0.010$ & $4.530 \pm 0.030$ \\
    $[\mathrm{Fe}/\mathrm{H}]$ (dex) & $-0.812 \pm 0.237$ & $-0.273 \pm 0.005$ & $0.260 \pm 0.080$ & $0.220 \pm 0.100$ \\
    $\pi_p$ (mas) & $73.270 \pm 0.340$ & $19.810 \pm 0.320$ & $747.170 \pm 0.610$ & $747.170 \pm 0.610$ \\
    \hline
    \multicolumn{5}{c}{\textbf{Photometric Data}} \\
    \hline
    $B$ (mag) & $4.620 \pm 0.007$ & $5.400 \pm 0.005$ & $0.720 \pm 0.100$ & $2.210 \pm 0.100$ \\
    $V$ (mag) & $4.120 \pm 0.008$ & $4.370 \pm 0.010$ & $0.010 \pm 0.100$ & $1.330 \pm 0.100$ \\
    $J$ (mag) & $3.300 \pm 0.286$ & $2.676 \pm 0.010$ & $-1.454 \pm 0.133$ & $-0.010 \pm 0.100$ \\
    $H$ (mag) & $2.988 \pm 0.240$ & $2.164 \pm 0.010$ & $-1.886 \pm 0.220$ & $-0.490 \pm 0.100$ \\
    $K$ (mag) & $4.640 \pm 0.010$ & $2.052 \pm 0.010$ & $-2.008 \pm 0.260$ & $-0.600 \pm 0.100$ \\
    $G$ (mag) & $4.000 \pm 0.001$ & $4.064 \pm 0.002$  & -- & -- \\
    $G_{\mathrm{BP}}$ (mag) & $4.261 \pm 0.001$ & $4.585 \pm 0.003$ & -- & -- \\
    $G_{\mathrm{RP}}$ (mag) & $3.556 \pm 0.002$ & $3.404 \pm 0.004$ & -- & -- \\
    \hline
    \multicolumn{5}{c}{\textbf{Spectroscopic Data}} \\
    \hline
    \shortstack{Date of \\ observation} & 21-08-2005 & 11-11-2014 & 06-05-2001 & 20-10-2020 \\
    Instrument & HARPS & HARPS & UVES & UVES \\
    \shortstack{No. of \\ observations} & 1 & 1 & 1 & 1 \\
    File name & \shortstack{ADP.2014-10-02\\T10\_02\_41.023.fits} &
    \shortstack{ADP.2014-11-11\\T19\_00\_32.193.fits} &
    \shortstack{ADP.2020-08-14\\T08\_30\_30.047.fits} &
    \shortstack{ADP.2020-08-04\\T17\_50\_24.992.fits} \\
    \hline
    \end{tabular}
    \tablefoot{The reference data for HD 222368 and HD 19787 are sourced from Gaia DR3 \citep{gaiadr3}. In the case of HD 128620 ($\alpha$ Cen A) and HD 128621 ($\alpha$ Cen B), where Gaia data are unavailable, the stellar parameters and abundance are taken from \citet{gent2022} and parallax from \citet{kervella2016}. The photometric values are compiled from various sources: for HD 222368, data is taken from \citet{Oja1993}, \citet{ducati}, \citet{2mass}, and \citet{vanbelle2009}; for HD 19787, the sources include \citet{ducati} and \citet{Laney2012}; and for HD 128620 and HD 128621, data is gathered from \citet{ducati} and \citet{2mass}, with uncertainties not available in the catalogue taken as $0.1$ mag. The spectroscopic data is obtained from HARPS \citep{Harps2003} and UVES \citep{Dekker2000}.}
    \label{tab:table3}
\end{table*}

\section{Results}
\label{sec:4}

\begin{table*}[h]
    \centering
    \caption{Estimation of the stellar parameters using the combined method.}
    \label{tab:table4}
    \small
    \begin{tabular}{lcccccccc}
    \hline
    \hline
    \noalign{\vskip 0.05cm}
     \shortstack[c]{Star} & \shortstack[c]{Correlation \\ Matrix} & \shortstack[c]{$T_{\text{eff}}$\\ (K)} & \shortstack[c]{$\log g$\\ (dex)} & \shortstack[c]{$\theta$\\ (mas)} & \shortstack[c]{Radius \\($R_{\odot}$)} & \shortstack[c]{Reduced\\ $\chi^2$} & \shortstack[c]{DoF} \\
    \hline
    \multirow{2}{*}{HD 222368 ($\iota$ Psc)} & \multirow{1}{*}{N}  & $6408 \pm 71$ & $4.419 \pm 0.186$ & $1.060 \pm 0.009$ & $1.555 \pm 0.015$ & \multirow{1}{*}{0.835} & \multirow{2}{*}{510} \\
     & \multirow{1}{*}{Y}  & $\mathbf{6403 \pm 64}$ & $\mathbf{4.426 \pm 0.328}$ & $\mathbf{1.081\pm 0.007}$ & $\mathbf{1.587 \pm 0.013}$ & \multirow{1}{*}{\textbf{2.454}} & \\
    \hline
     \multirow{2}{*}{HD 19787 ($\delta$ Ari)} & \multirow{1}{*}{N} & $4789 \pm 43$ & $2.858 \pm 0.218$ & $1.868 \pm 0.006$ & $10.136 \pm 0.164$ & \multirow{1}{*}{0.533} & \multirow{2}{*}{600} \\
     & \multirow{1}{*}{Y} & $\mathbf{4782 \pm 52}$ & $\mathbf{2.885 \pm 0.220}$ & $\mathbf{1.860 \pm 0.004}$ & $\mathbf{10.095 \pm 0.165}$ & \multirow{1}{*}{\textbf{0.897}} &  \\
    \hline
    \multirow{1}{*}{HD 128620 ($\alpha$ Cen A)} & \multirow{1}{*}{\textbf{N}}  & $\mathbf{5748 \pm 29}$ & $\mathbf{4.124 \pm 0.094}$ & $\mathbf{8.499 \pm 0.004}$ & $\mathbf{1.223 \pm 0.001}$ & \multirow{1}{*}{\textbf{2.296}} & \multirow{1}{*}{216} \\
      K17 &   & $5795 \pm 19$ & & $8.502 \pm 0.042$ & $1.224 \pm 0.006$ & $3.90$ & \\
      \hline
     \multirow{1}{*}{HD 128621 ($\alpha$ Cen B)} & \multirow{1}{*}{N} & $\mathbf{5154 \pm 26}$ & $\mathbf{4.541 \pm 0.034}$ & $\mathbf{6.001 \pm 0.003}$ & $\mathbf{0.864 \pm 0.001}$ & \multirow{1}{*}{\textbf{1.833}} & \multirow{1}{*}{285} \\
      K17 & & $5231 \pm 21$ & & $5.999 \pm 0.029$ & $0.863 \pm 0.005$ & $3.33$ & \\
    \hline
    \hline
    \end{tabular}
    \tablefoot{The table shows the results from the fitting method using the correlation matrix and without it. The uncertainty is presented as the full-width half maximum of the Gaussian distribution obtained from the MCMC analysis. The uncertainty associated with the estimated radius is determined using standard error propagation based on the parameters $\theta$ and $\pi_p$. For the $\alpha$ Cen targets, HD 128620 and HD 128621, we are comparing the results from the combined method to \citet{Kervella2017} (K17). The final results of our analysis are outlined in bold in the table.}

\end{table*}

Thanks to \texttt{PIMS}, we obtain high-precision estimations of the $\theta$, radius, and $T_{\textrm{eff}}$ for each science target. PIMS can also estimate $\logg$, but it is not well constrained by the interferometry data alone. The stellar parameters with their uncertainties and the $\chi^2_\textrm{red}$ values of the best model are highlighted in bold in Table \ref{tab:table4}. For the CHARA targets, we present the results without and with consideration of the correlation in the interferometric data.

\begin{figure*}
    \centering
    \includegraphics[width = 17 cm]{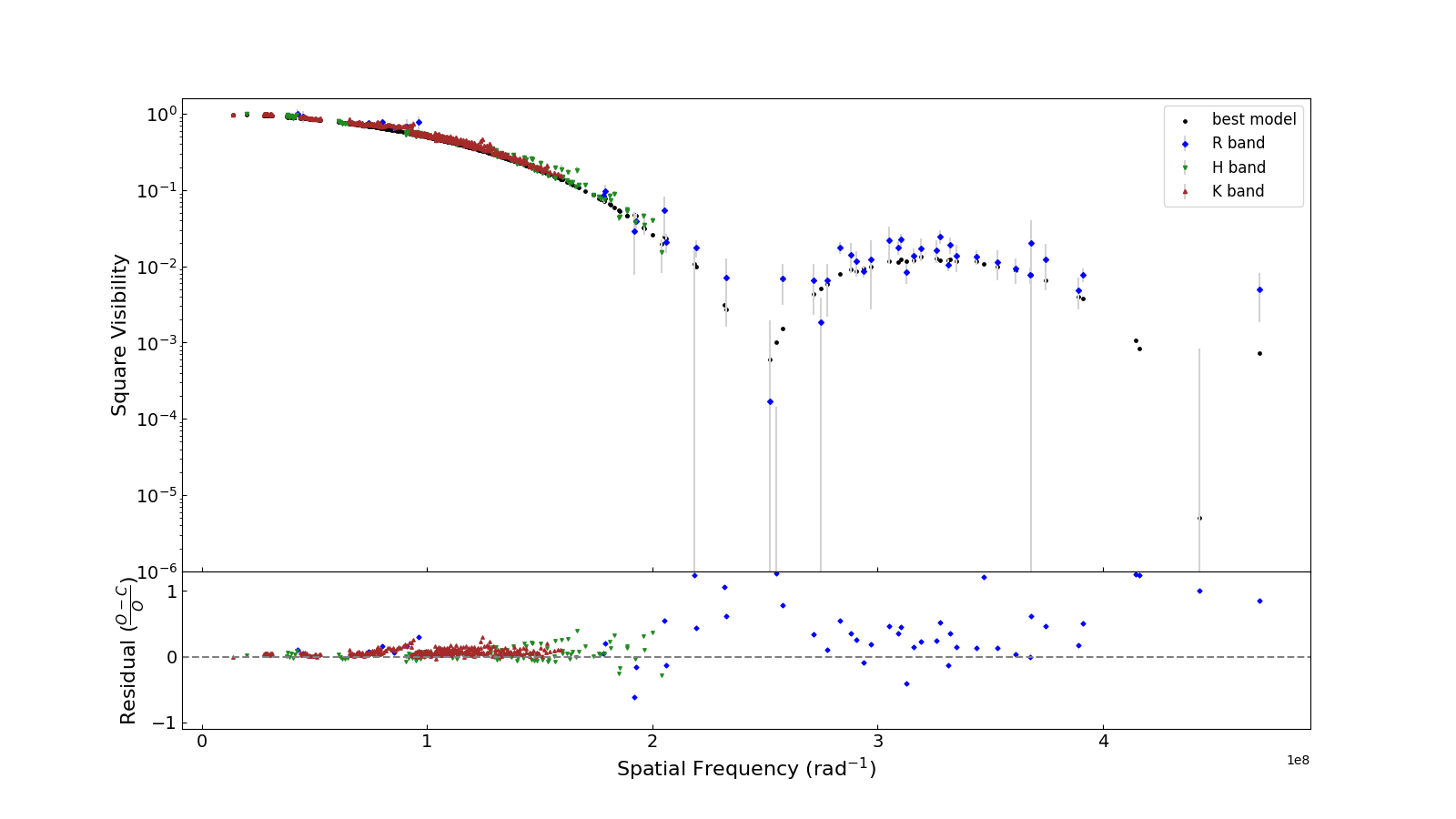}
    
    \caption{Data and the best-fitted model for $\iota$ Psc (HD~222368), considering the correlation in the data. (Top) The black points represent the best models in the $R$, $H$, and $K$ bands from the ANN trained on MARCS models. The dark blue, green, and maroon points are the observations from SPICA ($R$ band), MIRC-X ($H$ band), and MYSTIC ($K$ band), while the grey lines indicate the observational uncertainties. (Bottom) Ratio of the residuals between the observed square visibilities and the modelled square visibilities in different bands of observations. The grey horizontal dashed line shows the zero residual. }
    \label{fig:fig2}
\end{figure*}

\begin{figure*}
    \centering
    \includegraphics[width = 17 cm]{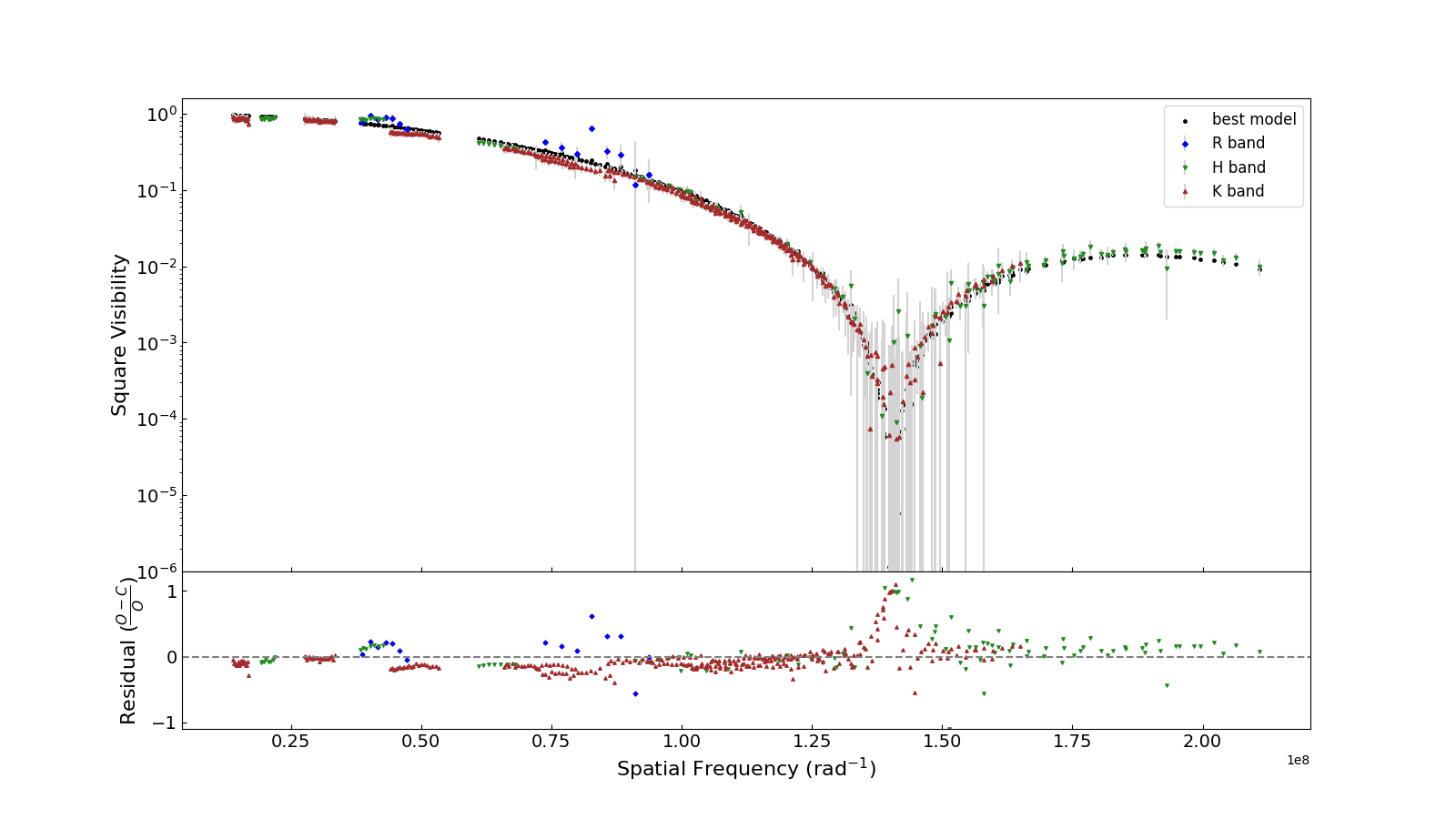}
    
    \caption{Data and the best-fitted model for $\delta$ Ari (HD~19787), considering the correlation in the datasimilar to Figure~\ref{fig:fig2}. }
    \label{fig:fig3}
\end{figure*}

\begin{figure}[ht!]
    \centering

    \begin{subfigure}[t]{0.45\textwidth}
        \centering
        \includegraphics[width=\textwidth]{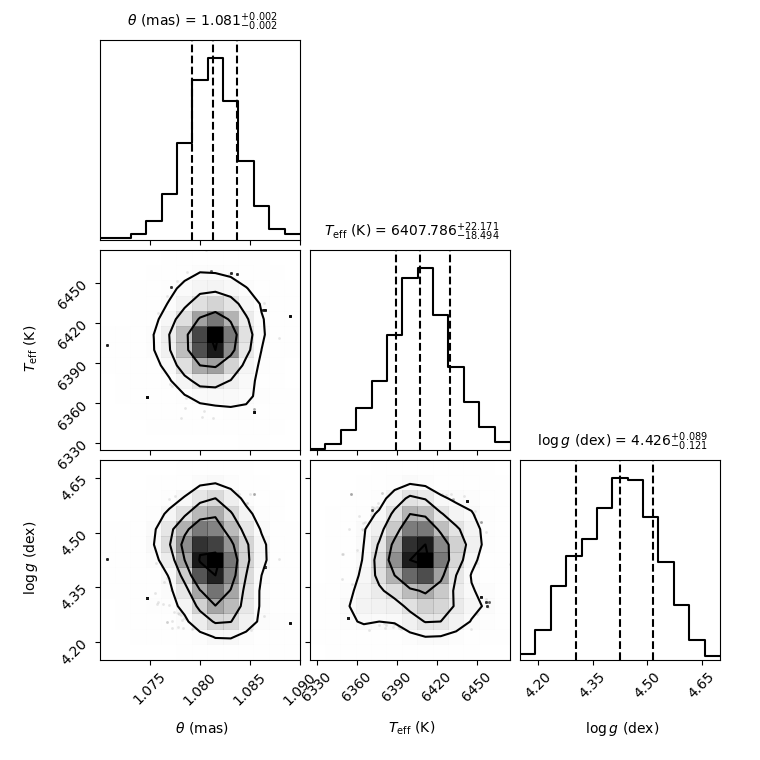}
        
    \end{subfigure}
    \hfill

    \begin{subfigure}[t]{0.45\textwidth}
        \centering
        \includegraphics[width=\textwidth]{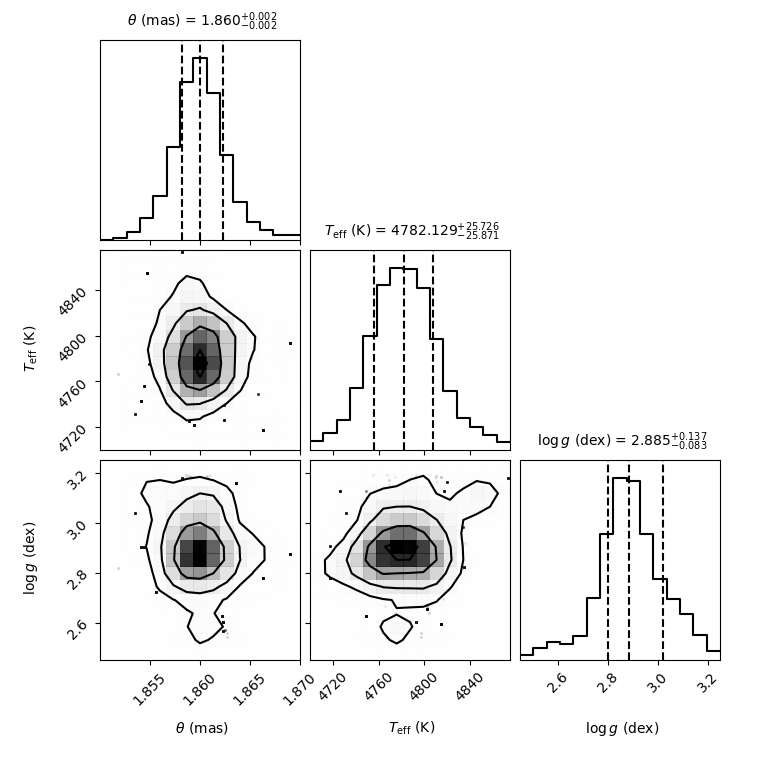}
        
    \end{subfigure}    
\caption{The corner plot for the $\iota$ Psc (HD~222368, top) and $\delta$ Ari (HD~19787, bottom) with the correlation between different baselines of the observations for both calibrators and targets throughout the night. The contours represent the $1\sigma$, $2\sigma$ and $3\sigma$ of the distribution.}
\label{fig:fig4}
\end{figure}

Figure \ref{fig:fig2} presents the data and the best model, including correlation, for the target HD 222368 and Figure~\ref{fig:fig3} shows the corresponding fit of HD 19787. The black points represent the best-fitting models in the H, K, and R bands from the ANN, while the dark blue, green, and maroon points correspond to the observations from SPICA (R band), MIRC-X (H band), and MYSTIC (K band), respectively. The good visual agreement across the different bands is supported by the final $\chi^2_\textrm{red}$ values. 
Figure~\ref{fig:fig4} presents the corner plot of the probability distribution of the parameters using the MCMC sampler for HD~222368 (top) and HD 19787 (bottom). Similar figures showing the model-fitting results for targets without the correlation matrix and $\alpha$ Cen targets are presented in the Appendix~\ref{sec:AppA}.

\subsection{$\chi^2$ analysis - parameters and uncertainties}
\label{sec:4.1}

In Sect.~\ref{sec:2.4}, we presented how the correlation existing in the observations is used to improve the reliability of the $\chi^2_\textrm{red}$ statistics. In Table~\ref{tab:table4}, we present the values of the parameters estimated in both scenarios, with and without considering the correlation. The first conclusion is that, in both cases, the parameters are within the limits of the uncertainties of the initial estimations, such as those provided by \texttt{JMMC GetStar}. However, the small $\chi^2_\textrm{red}$ values ($0.835$ and $0.533$ for HD~222368 and HD~19787, respectively) without considering the correlation indicate that the uncertainties on the data are probably overestimated.
In contrast, when the correlation is taken into account, the $\chi^2_\textrm{red}$ values are increased ($2.454$ and $0.897$ for HD~222368 and HD~19787, respectively), indicating a more reliable estimation of the parameters. 

\begin{table*}[ht!]
\centering
\caption{Reduced $\chi^2$ for the observations from ISSP.}
\small
\begin{tabular}{c  c  cc  cc  cc}
\hline
\multirow{2}{*}{} 
& \multirow{2}{*}{Star} 
& \multicolumn{2}{c}{SPICA} 
& \multicolumn{2}{c}{MIRC-X} 
& \multicolumn{2}{c}{MYSTIC} \\
&
& $\chi^{2}_{\mathrm{red}}$ & dof
& $\chi^{2}_{\mathrm{red}}$ & dof 
& $\chi^{2}_{\mathrm{red}}$ & dof \\
\noalign{\vskip 0.051cm}
\hline
\multirow{2}{*}{Without correlation} & HD 222368 & $0.533$ & $75$ & $1.264$ & $120$ & $0.747$ & $315 $\\
& HD 19787 & $0.842$ & $165$ & $0.585$ & $120 $ & $0.352$ & $315$ \\
\hline
\multirow{2}{*}{Correlation} & HD 222368 & $0.492$ & $75$ & $2.562$ & $120$ & $2.879$ & $315$ \\
& HD 19787 & $0.887$ & $165$ & $0.828$ & $120$ & $0.925$ & $315$ \\
\hline

\end{tabular}
\tablefoot{The table presents the $\chi^2_\textrm{red}$ values for data from different instruments for ISSP observations, along with the degrees of freedom (dof) for each observation.}
\label{tab:table5}
\end{table*}

A closer inspection of the $\chi^2_\textrm{red}$ values for the individual datasets from the ISSP observations (see Table~\ref{tab:table5}) shows similar overall behavior. In most cases, the $\chi^2_\textrm{red}$ values indicate consistent fits. However, for the SPICA observations of HD 19787, the $\chi^2_\textrm{red}$ exhibits little change between the correlated and uncorrelated fits. This outcome is expected because the dataset includes only two baselines from the first observing night, which limits our ability to constrain the fit meaningfully. 

In the case of HD~222368, in MIRC-X and MYSTIC data, the $\chi^2_\textrm{red}$ increases when the correlation matrix is applied. This increase is due to the propagation of the correlated uncertainties in the square visibilities. For the SPICA data, no significant change in $\chi^2_\textrm{red}$ is observed when correlations are included. This is attributed to the smaller number of data points and to the rescaling of the visibility when applying the correction of the calibration by the background component.

In addition, we estimate the uncertainties of each parameter using the chi-square plus one ($\chi^2 + 1$) criterion as an independent validation method. The $\chi^2 + 1$ criterion determines the 1-$\sigma$ uncertainty of a parameter value at the minimum value of $\chi^2$. This is done by identifying the range of that parameter where $\chi^2$ increases by 1 above its minimum. The results obtained using this approach, calculated for each star both with and without the correlation matrix, are consistent with those from the MCMC analysis, in Table \ref{tab:table4}, further supporting the validity of our parameter estimates.

\section{Discussion}

Using PIMS, we also fitted the interferometric observations of $\alpha$~Cen presented by \citet{Kervella2017}. We then investigated the implications of the parameters estimated using PIMS for the precise determination of stellar masses and ages. These are discussed in the following two subsections.

\subsection{$\alpha$ Centauri data}
\label{sec:4.2}
\citet{Kervella2017} present interferometric measurements of the LD angular diameters of $\alpha$ Cen A and B from VLTI/PIONIER with an accuracy of $0.4 \%$. They conducted a model fitting of the data using different LD laws. Among these, the model with the lowest reduced $\chi^2_\textrm{red}$ was the power-law LD model, which yielded a $\chi^2_\textrm{red}$ of $3.90$ for $\alpha$~Cen A and $3.33$ for $\alpha$ Cen B. Using a square-root LD law, they got $\chi^2_\textrm{red}$ of $4.24$ for $\alpha$ Cen A.  For $\alpha$ Cen B, there was no fit with the square-root LD law. The results of the analysis of these data sets with our new method are presented in Table~\ref{tab:table4}. We obtain a $\chi^2_\textrm{red}$ value of $2.296$ for $\alpha$ Cen A and $1.833$ for $\alpha$ Cen B. The $\theta$ values ($8.499 \pm 0.004$ mas and $6.001 \pm 0.003$ mas for A and B, respectively) are within the $1\sigma$ range of the values estimated by \citet{Kervella2017}, which were $8.502 \pm 0.042$ mas for $\alpha$ Cen A and $5.999 \pm 0.029$ mas for $\alpha$ Cen B. However, one can note that their $\theta$ value for $\alpha$ Cen A, when using the square-root law ($8.446 \pm 0.017$ mas), is at almost  $4\sigma$ from our estimation. The improvement in the $\chi^2_\textrm{red}$ value from the combined method could explain this discrepancy. \citet{Kervella2017} calculated radii of $1.224 \pm 0.006$ $R_\odot$ for $\alpha$~Cen~A and $0.863 \pm 0.005$ $R_\odot$ for $\alpha$~Cen~B. 

They also determined $\teff$ of $5795 \pm 19$ K for $\alpha$ Cen A and $5231 \pm 21$ K for $\alpha$ Cen B. The $\teff$ and radii derived from our combined fitting method (see Table~\ref{tab:table4}) are consistent with these values within the $1\sigma$ uncertainties for both stars. This consistency demonstrates the robustness and reliability of the combined method, highlighting its ability to recover precise and accurate stellar parameters by jointly exploiting interferometry, spectroscopy, and photometry. It should be noted, however, that in the case of these VLTI data, we do not have access to the information required to estimate the correlation in the data, and we have only one available spectral band, reducing our ability to identify possible bias. The uncertainties obtained on the parameters are thus probably under-evaluated.

As an additional check, we also applied the $\chi^2 + 1$ uncertainty estimation method to the same datasets. The parameter values obtained using this approach likewise remain within the $1\sigma$ ranges of both the \citet{Kervella2017} measurements and our combined-method results, further confirming the robustness and reliability of our analysis.

\subsection{Application: Impact on mass and age}
\label{sec:6}

In this section we investigate the impact of our new parameters on the  determination of mass and age. To do this we employ the SPInS code \citep{spins,spinscode}, which is a parameter optimisation method based on a MCMC approach, exploiting stellar evolution models in order to derive masses and ages of stars.  We use the latest version from \citet{spinsupdate}, which is the code that has also been implemented in the Gaia astrophysical parameters software code, Apsis/FLAME \citep{flame}, and will produce results in Gaia Data Release 4 and 5.  We also use the same BASTI stellar evolution models \citep{basti2018} that are implemented in Apsis/FLAME.  The grid spans from 0.1 to 10 M$_\odot$, from --3.7 to +0.45 dex in metallicity, and with ages from the zero-age-main-sequence (defined here as 0 Ma) to the tip of the red giant branch.   

SPInS takes as input a variety of observational constraints.  For our use here, we coupled the derived radius, temperature, and $\log g$ from this paper with metallicities and luminosities from the literature. We take a different approach for the metallicities compared to the first part of the work, where we aimed to follow a homogeneous approach. However, since the evolution models are sensitive to metallicity, we adopted more precise abundance values here. 
The metallicities, for HD\,222368 and HD\,19787, have been taken from \cite{gbsv3mh} while those for the $\alpha$ Cen system are from \cite{Jofre2014}. The luminosities for HD 222368 and HD 19787 were obtained from Gaia DR3 \citep{gaiadr3}, and the luminosities for the $\alpha$ Cen system are from \citet{heiter2015}.  The full list of constraints are shown in Table~\ref{tab:table6}.

As can be seen in Table~\ref{tab:table6}, for HD\,222368 we have different observational constraints. 
Before deriving the optimal stellar parameters for the four stars studied in this paper, we use this star to first discuss the impact of the different observational constraints on the model stellar parameters.

\begin{table*}[]
    \centering
    \caption{Mass and age determination }
    \begin{tabular}{llllllll}
    \hline\hline
star & case &  \teff\ & $\cal{R}$& $\log g$ & $\cal{L}$&   [Fe/H]  \\
& & [K] & [R$_\odot$] & [dex] & [L$_\odot$] & [dex]\\
    \hline
    HD~222368& 1 & 6403 $\pm$ 64 & 1.587 $\pm$ 0.013 & 4.426 $\pm$ 0.328 & 3.554 $\pm$ 0.061&  --0.11 $\pm$ 0.06$^a$ \\
     HD~222368 & 2 & 6403 $\pm$ 64 & 1.587 $\pm$ 0.013 & 4.426 $\pm$ 0.328 & 3.554 $\pm$ 0.061& --0.14 $\pm$ 0.02$^b$ & \\
     HD~222368 & 3 & 6403 $\pm$ 64 & 1.587 $\pm$ 0.013 & 4.426 $\pm$ 0.328 & 3.554 $\pm$ 0.061 & --0.08 $\pm$ 0.02$^c$ & \\
     HD~222368& 4 & 6403 $\pm$ 64 & -- & 4.426 $\pm$ 0.328 & 3.554 $\pm$ 0.061 & --0.11 $\pm$ 0.06 &\\
    HD~222368 & 5 & 6403 $\pm$ 120 & -- & 4.426 $\pm$ 0.328 & 3.554 $\pm$ 0.061 & --0.11 $\pm$ 0.06 &\\
    HD~19787 & 6  & 4782 $\pm $ 52 & 10.095 $\pm$ 0.163 & 2.885 $\pm$ 0.220 & 52.7 $\pm$ 1.6 & --0.096 $\pm$ 0.05$^d$ \\
    HD~128620 & 7& 5748 $\pm$ 29 & 1.223 $\pm$ 0.001 & 4.124 $\pm$ 0.094 & 1.52 $\pm$ 0.01 & 0.26 $\pm$ 0.08\\
    HD~128621 & 8& 5154 $\pm$ 26 & 0.864 $\pm$ 0.001 & 4.541 $\pm$ 0.034 & 0.50 $\pm$ 0.01 & 0.22 $\pm$ 0.10\\
    
         \hline\hline

    \end{tabular}
    \label{tab:table6}
    \tablefoot{ Observational constraints for the mass and age determination for the four stars in this study.  The column heading `case' is used within the text in this section to help identify which observational constraints the results correspond to. $^a$Using an average of FeI and FeII lines. $^b$Using FeI lines. $^c$Using FeII lines. $^d${FeI lines}. }
\end{table*}

\subsubsection{Impact of metallicity}
\label{sec:6.1}
\begin{figure}
    \centering
    \includegraphics[width=0.95\linewidth]{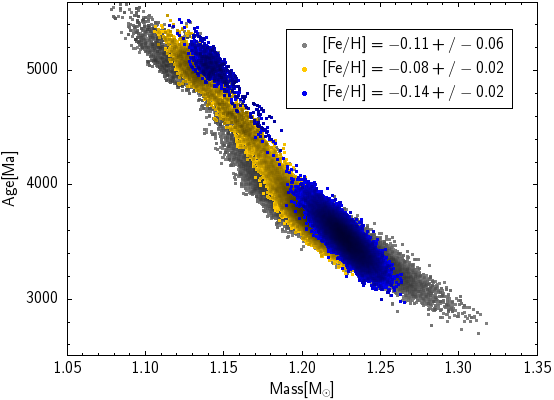}
    \caption{Distribution of the MCMC samples of the mass and age parameters for $\iota$ Psc (HD\,222368).  The different colours represent the results when we assume different metallicities.}
    \label{fig:fig5}
\end{figure}

We begin by illustrating the impact of the metallicity parameter on the final model results for the star \object{HD\,222368}.  Stellar evolution is most sensitive to mass and chemical composition, and therefore the input metallicity of the star will have an impact on the derived mass and age.  
In Figure~\ref{fig:fig5} we show this for cases 1, 2, and 3 from Table~\ref{tab:table6}, where we compared the effect of using the average abundance of Fe~I and Fe~II with the results obtained when considering them separately. This figure shows the distribution of the output MCMC samples from which the mass and age are derived.  
The grey points correspond to the solution when the mean metallicity value with the larger uncertainty is used (case 1), while the blue and the yellow points correspond to the results when using a lower and higher metallicity.  It is very clear from this figure that the metallicity parameter plays a decisive role in the resulting mass and age.  
For the three cases, the derived masses are $1.21 \pm 0.04$, $1.20 \pm 0.03$, and $1.23 \pm 0.03$ M$_\odot$, while the corresponding ages are $3.67 \pm 0.55$, $3.75 \pm 0.47$, $3.55 \pm 0.45$ Ga.   In this particular case, as the metallicities are quite similar, the differences are not significant.   

\begin{figure}
    \centering
    \includegraphics[width=0.95\linewidth]{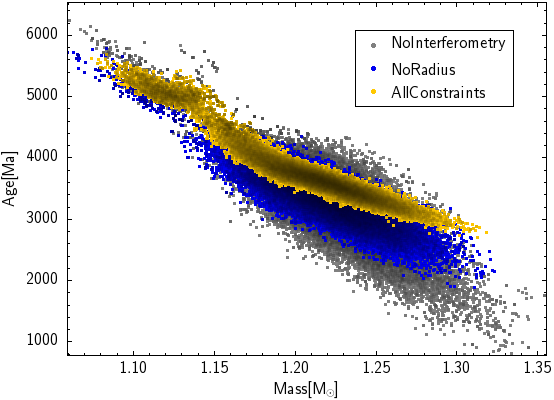}
    \caption{Distribution of the MCMC samples of the mass and age parameters for $\iota$ Psc (HD\,222368).  The different colours represent the results when we use different observational constraints. The distribution shown in this figure, with all constraints (in yellow ), corresponds to [Fe/H] = $-0.11 \pm 0.06$ shown in Figure~\ref{fig:fig5} by the grey points.}
    \label{fig:fig6}
\end{figure}

\subsubsection{Impact of interferometry}
\label{sec:6.2}
The inclusion of interferometric constraints brings two main additions to the posed problem: firstly the uncertainties in the \teff\ decrease typically to less than 50 K for well determined diameters, which is a reduction by a factor of two in the typical uncertainty from \teff\ spectroscopic measurements.  
Secondly, although the radius and \teff\ are correlated if they are derived using the angular diameter, the addition of the radius as a constraint contributes to reducing the correlations.  Indeed if we use only luminosity and \teff\ we are neglecting some information.

In Figure~\ref{fig:fig6} we show the distribution of the MCMC samples for cases 1, 4 and 5.  
The grey points correspond to the solution when we have no interferometric constraints, i.e. no radius measurement, and an increase in the \teff\ error to 120 K.  The blue points represent the results when we have an interferometric diameter but we only include the \teff\ as an observational constraint, and the yellow represents case 1 when we also include the radius as a constraint.  

The reported values from the different cases are all in agreement in mass and age and their 1D uncertainties do not change by much, however, one can clearly see that there is a much tighter constraint in the parameter space of mass and age, where there are clear areas that the solution would not be allowed due to the constraint on the interferometry.

We apply the methodology to the four stars in this study and we report their masses and ages  in Table~\ref{tab:table7}.  
For HD\,222368 we consider case 1 as the reference case.  
For the $\alpha$ Cen A and B we report their masses and ages individually, and then we re-analyse the two stars as a system by imposing the same age and metallicity.    

\paragraph{\textbf{HD 222368:}}
\citet{Allende1999} estimated a stellar mass of $1.38 \pm 0.07 $ \msol\ for HD 222368 using stellar evolution models, while the TESS Input Catalogue (TIC; \citealt{Paegert2021}) reports $1.21 \pm 0.17 $ \msol. In our analysis, using the latest observational constraints available for this star, we obtain a mass in close agreement with the latter but with much higher precision; $1.21 \pm 0.04$ \msol.  This results in an age that is well constrained to $3.7 \pm 0.6$ Ga.  

\paragraph{\textbf{HD 19787:}}
For HD 19787, \citet{Allende1999} derived a mass of $1.50 \pm 0.65 M_\odot$, while \citet{Hekker2006} obtained $1.91 M_\odot$ using a similar method.  With our new observational constraints we were able to refine this value to $1.89 \pm 0.18$ \msol, resulting in an age of $1.3 \pm 0.4$ Ga.

\begin{figure}
    \centering
    \includegraphics[width=0.95\linewidth]{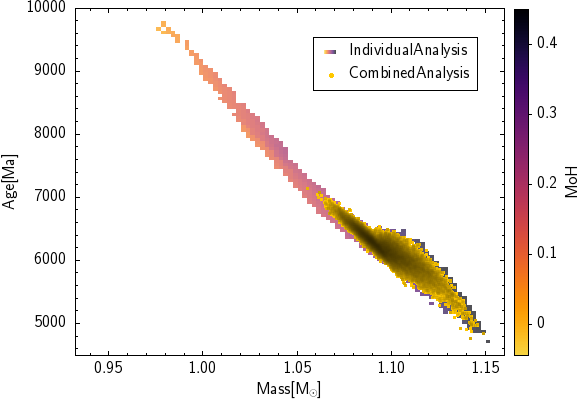}
    \includegraphics[width=0.95\linewidth]{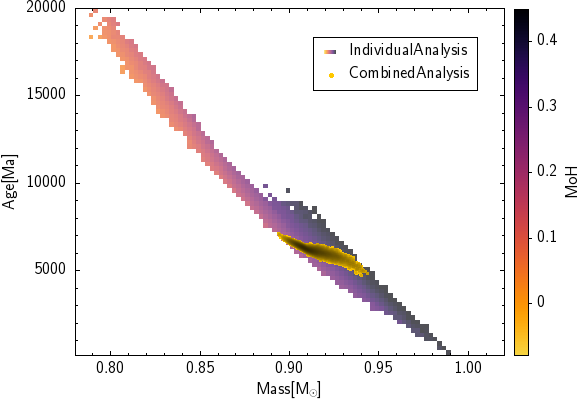}
    \caption{Distribution of the MCMC samples for mass and age for $\alpha$ Cen A (top) and B (bottom).  The background histogram plot, colour-coded by metallicity (written as `MoH'), represents the samples when the stars are analysed individually.  The foreground yellow plot are the samples when the two stars are analysed as one system and the same age is imposed. 
    }
    \label{fig:fig7}
\end{figure}

\paragraph{\textbf{$\alpha$ Cen A and B:}} 
The distribution of the masses and ages for $\alpha$ Cen A and B from the MCMC samples are shown  in Figure~\ref{fig:fig7} by the background histogram plot.  The top panel represents the distributions for $\alpha$ Cen A while the lower shows $\alpha$ Cen B. As one can see, without imposing an age prior on $\alpha$~Cen~B, its resulting age can reach values beyond the age of the universe. These older values do however also correspond to lower metallicity values, as indicated by the colour-code, which tend to diverge from the typical measured values.  

However, as this is a co-eval system, we could also model the two stars by imposing that they have the same age.   When we do this, we report extremely high precision results for both components of the system.  The resulting distributions in mass and age are shown by the yellow density plot imposed on the background histogram plot in Figure~\ref{fig:fig7}.  The resulting age is 6.22 $\pm$ 0.22 Ga, corresponding to a mass of $1.094 \pm 0.011$ M$_\odot$ for the primary and $0.911 \pm 0.007$ M$_\odot$ for the secondary. 

Our analysis of the components of $\alpha$ Cen results in masses and ages in good agreement with the literature.  For example, \cite{akeson21} reports $1.0788 \pm 0.0029$ \msol\ for A and $0.9092 \pm 0.0025$ \msol\ for B component using ALMA high precision astrometry for the system.
\cite{gbsv3} uses a similar approach to the one taken in this paper by comparing observational data to stellar evolution models, and they report a mass of 1.04 or 1.08 \msol\ for A and 0.86 or 0.96 \msol\ for B, where the two estimates are based on the use of different stellar evolution models.
Our results for component B fall right between the two estimates and are in excellent agreement with the ALMA astrometric measurements.  
For component A we obtain a result in statistical agreement but on the slightly higher end range.  
Earlier works had also reported slightly higher masses, such as \cite{thevenin2002} or \cite{Kervella2017} who both report a mass of 1.10 \msol.

Concerning the system age, several studies of the last two decades have estimated the age to be between 4 and 8 Ga, for example \citet{thevenin2026} provide an age estimate of the system of between $7.8 \pm 0.6$ Ga and $8.7 \pm 0.6$ Ga depending on the initial metallicity, while \citet{joyce2018} provide an estimate of $5.3 \pm 0.3$ Ga.
Our joint analysis ($6.22\pm 0.22$\,Ga) 
 results in an age compatible on the absolute scale, however the differences are typically larger than 3$\sigma$ considering only our uncertainties.   The absolute differences are in part due to the assumptions of the physics in the models and the initial metallicity, as well as the input observations.  The $\sigma$-differences, however, reflect the fact that our uncertainties are based on models with only three free parameters; the BASTI public models impose a chemical enrichment law, which fixes the initial composition, and the mixing-length parameter is fixed to the one based on a solar-calibrated model. 

\begin{table}[]
    \centering
    \caption{Derived masses and ages of the four studied stars.}
    \begin{tabular}{lcccc}
    \hline\hline
    Star & Mass & Age  \\ 
    & [M$\odot$] & [Ga] \\
    \hline
    HD\,222368 & $1.21 \pm 0.04$ & $3.7 \pm 0.6$\\
    HD\,19787 & $1.89 \pm 0.18$& $1.3 \pm 0.4$\\
    HD\,128620 & $1.08 \pm 0.03$ & $6.5 \pm 0.6$ \\
    HD\,128621 & $0.90 \pm 0.03$ & $7.4 \pm 3.2$ \\
        HD\,128620$_{\rm comb}$ & $1.094 \pm 0.011$ & $6.22 \pm 0.22$ \\
        HD\,128621$_{\rm comb}$ & $0.911 \pm 0.007$ & $6.22 \pm 0.22$ \\
         \hline\hline
    \end{tabular}
    
    \label{tab:table7}
\end{table}

\section{Conclusion}
\label{sec:7}

We demonstrated how we can estimate the angular diameter and other stellar parameters employing an ANN that integrates stellar atmosphere models and multiple observational data,  interferometry, spectroscopy, and photometry. This methodology combined with multi-band interferometric data permits us to estimated stellar parameters with an increased precision. The PIMS was demonstrated for both a dwarf star ($\iota$ Psc) and a giant star ($\delta$ Ari) from the ongoing ISSP observations. We have also shown how the existing correlation in the interferometric data can be taken into account to improve the reliability of the adjustment. The estimated angular diameters and stellar parameters using PIMS and the correlation matrix for these stars are highlighted in bold in Table~\ref{tab:table4}. 

Fitting the VLTI/PIONIER data for $\alpha$ Cen A and B helps in consolidating the robustness of our method. We showed that our method can improve the accuracy of the fit, with respect to the work published in \citet{Kervella2017}, as indicated by the $\chi^2_\textrm{red}$ and the parameter values obtained for these targets. With the resulting parameters from this work we also estimated the masses and ages of the four target stars, and our new results are consistent with previous literature results while providing higher precision in most cases.

The ISSP aims to conduct a large-scale interferometric survey of hundreds of stars across the HR diagram. One of the objectives is the measurement of angular diameter and limb darkening from interferometric observations to estimate stellar fundamental parameters using data from the SPICA, MIRC-X, and MYSTIC instruments. Over the past three years of observations within the ISSP, we have observed about a hundred stars for this science goal, with several stars having multiple observations. Given the accuracy attainable through this method, we will be able to derive reliable parameters for these targets. The interferometric data and estimated stellar parameters will be made publicly available through OiDB and other catalogues. The results will be published in a forthcoming paper.

Finally, we have integrated the interferometric module in the PLATO MSteSci1 pipeline. This will enable us to explore a wide range of stars with interferometric observations from various instruments in different spectral bands and resolutions.

\begin{acknowledgements}
This project has received funding from the European Research Council (ERC) under the European Union’s Horizon 2020 research and innovation programme (Grant agreement No. 101019653). We acknowledge the ISSP team for providing the observations used in this work. This research has made use of the Jean-Marie Mariotti Center at https://jmmc.fr/. This research has made use of the SIMBAD database, operated at CDS, Strasbourg, France. This work is based upon observations obtained with the Georgia State University Center for High Angular Resolution Astronomy Array at Mount Wilson Observatory.  The CHARA Array is supported by the National Science Foundation under Grant No. AST-2034336 and AST-2407956. Institutional support has been provided from the GSU College of Arts and Sciences, Office of the Provost, and Office of the Vice President for Research and Economic Development. SK acknowledges support from an ERC Consolidator Grant (Grant Agreement ID 101003096). MIRC-X has been build with funds from an ERC Starting Grant (Grant Agreement No.\ 639889) and an STFC equipment grant ST/X005143/1. JDM acknowledges funding for the development of MIRC-X (NASA-XRP NNX16AD43G, NSF-AST 2009489) and MYSTIC (NSF-ATI 1506540, NSF-AST 1909165). TM acknowledges financial support from Belspo for contract PRODEX PLATO mission development. BP is supported in part by the french space agency CNES. 

We warmly thank Chris Farrington, Becky Flores, Olli Majoinen, Heven Renteria, and Norm Vargas for their excellent support during the night operations.
\end{acknowledgements}

\bibliographystyle{aa}
\bibliography{refer}

@ARTICLE{Gustafsson2008,
       author = {{Gustafsson}, B. and {Edvardsson}, B. and {Eriksson}, K. and {J{\o}rgensen}, U.~G. and {Nordlund}, {\r{A}}. and {Plez}, B.},
        title = "{A grid of MARCS model atmospheres for late-type stars. I. Methods and general properties}",
      journal = {\aap},
         year = 2008,
        month = aug,
       volume = {486},
       number = {3},
        pages = {951-970},
          doi = {10.1051/0004-6361:200809724},
archivePrefix = {arXiv},
       eprint = {0805.0554},
 primaryClass = {astro-ph},
       adsurl = {https://ui.adsabs.harvard.edu/abs/2008A&A...486..951G}
}

@ARTICLE{Brown1974,
       author = {{Hanbury Brown}, R. and {Davis}, J. and {Lake}, R.~J.~W. and {Thompson}, R.~J.},
        title = "{The effects of limb darkening on measurements of angular size with an intensity interferometer}",
      journal = {\mnras},
         year = 1974,
        month = jun,
       volume = {167},
        pages = {475-484},
          doi = {10.1093/mnras/167.3.475},
       adsurl = {https://ui.adsabs.harvard.edu/abs/1974MNRAS.167..475H}
}

@ARTICLE{Kervella2017,
       author = {{Kervella}, P. and {Bigot}, L. and {Gallenne}, A. and {Th{\'e}venin}, F.},
        title = "{The radii and limb darkenings of {\ensuremath{\alpha}} Centauri A and B . Interferometric measurements with VLTI/PIONIER}",
      journal = {\aap},
         year = 2017,
        month = jan,
       volume = {597},
          eid = {A137},
        pages = {A137},
          doi = {10.1051/0004-6361/201629505},
archivePrefix = {arXiv},
       eprint = {1610.06185},
 primaryClass = {astro-ph.SR},
       adsurl = {https://ui.adsabs.harvard.edu/abs/2017A&A...597A.137K}
}

@ARTICLE{thevenin2026,
       author = {{Th{\'e}venin}, F. and {Baturin}, V.~A. and {Oreshina}, A.~V. and {Morel}, P. and {Ayukov}, S.~V. and {Bigot}, L. and {Gorshkov}, A.~B.},
        title = "{A mapping method of age estimation for binary stars: Application to the $α$ Centauri system A and B}",
      journal = {arXiv e-prints},
         year = 2026,
        month = feb,
          eid = {arXiv:2602.08879},
        pages = {arXiv:2602.08879},
          doi = {10.48550/arXiv.2602.08879},
archivePrefix = {arXiv},
       eprint = {2602.08879},
 primaryClass = {astro-ph.SR},
       adsurl = {https://ui.adsabs.harvard.edu/abs/2026arXiv260208879T}
}

@ARTICLE{kervella2016,
       author = {{Kervella}, P. and {Mignard}, F. and {M{\'e}rand}, A. and {Th{\'e}venin}, F.},
        title = "{Close stellar conjunctions of {\ensuremath{\alpha}} Centauri A and B until 2050 . An m$_{K}$ = 7.8 star may enter the Einstein ring of {\ensuremath{\alpha}} Cen A in 2028}",
      journal = {\aap},
         year = 2016,
        month = oct,
       volume = {594},
          eid = {A107},
        pages = {A107},
          doi = {10.1051/0004-6361/201629201},
archivePrefix = {arXiv},
       eprint = {1610.06079},
 primaryClass = {astro-ph.SR},
       adsurl = {https://ui.adsabs.harvard.edu/abs/2016A&A...594A.107K}
}

@ARTICLE{spins,
       author = {{Lebreton}, Y. and {Reese}, D.~R.},
        title = "{SPInS, a pipeline for massive stellar parameter inference. A public Python tool to age-date, weigh, size up stars, and more}",
      journal = {\aap},
         year = 2020,
        month = oct,
       volume = {642},
          eid = {A88},
        pages = {A88},
          doi = {10.1051/0004-6361/202038602},
archivePrefix = {arXiv},
       eprint = {2009.00037},
 primaryClass = {astro-ph.SR},
       adsurl = {https://ui.adsabs.harvard.edu/abs/2020A&A...642A..88L}
}

@ARTICLE{gbsv3,
       author = {{Soubiran}, C. and {Creevey}, O.~L. and {Lagarde}, N. and {Brouillet}, N. and {Jofr{\'e}}, P. and {Casamiquela}, L. and {Heiter}, U. and {Aguilera-G{\'o}mez}, C. and {Vitali}, S. and {Worley}, C. and {de Brito Silva}, D.},
        title = "{Gaia FGK benchmark stars: Fundamental T$_{eff}$ and log g of the third version}",
      journal = {\aap},
         year = 2024,
        month = feb,
       volume = {682},
          eid = {A145},
        pages = {A145},
          doi = {10.1051/0004-6361/202347136},
archivePrefix = {arXiv},
       eprint = {2310.11302},
 primaryClass = {astro-ph.SR},
       adsurl = {https://ui.adsabs.harvard.edu/abs/2024A&A...682A.145S}
}

@ARTICLE{joyce2018,
       author = {{Joyce}, M. and {Chaboyer}, B.},
        title = "{Classically and Asteroseismically Constrained 1D Stellar Evolution Models of {\ensuremath{\alpha}} Centauri A and B Using Empirical Mixing Length Calibrations}",
      journal = {\apj},
         year = 2018,
        month = sep,
       volume = {864},
       number = {1},
          eid = {99},
        pages = {99},
          doi = {10.3847/1538-4357/aad464},
archivePrefix = {arXiv},
       eprint = {1806.07567},
 primaryClass = {astro-ph.SR},
       adsurl = {https://ui.adsabs.harvard.edu/abs/2018ApJ...864...99J}
}

@ARTICLE{thevenin2002,
       author = {{Th{\'e}venin}, F. and {Provost}, J. and {Morel}, P. and {Berthomieu}, G. and {Bouchy}, F. and {Carrier}, F.},
        title = "{Asteroseismology and calibration of alpha Cen binary system}",
      journal = {\aap},
         year = 2002,
        month = sep,
       volume = {392},
        pages = {L9-L12},
          doi = {10.1051/0004-6361:20021074},
archivePrefix = {arXiv},
       eprint = {astro-ph/0206283},
 primaryClass = {astro-ph},
       adsurl = {https://ui.adsabs.harvard.edu/abs/2002A&A...392L...9T}
}

@ARTICLE{akeson21,
       author = {{Akeson}, Rachel and {Beichman}, Charles and {Kervella}, Pierre and {Fomalont}, Edward and {Benedict}, G. Fritz},
        title = "{Precision Millimeter Astrometry of the {\ensuremath{\alpha}} Centauri AB System}",
      journal = {\aj},
         year = 2021,
        month = jul,
       volume = {162},
       number = {1},
          eid = {14},
        pages = {14},
          doi = {10.3847/1538-3881/abfaff},
archivePrefix = {arXiv},
       eprint = {2104.10086},
 primaryClass = {astro-ph.SR},
       adsurl = {https://ui.adsabs.harvard.edu/abs/2021AJ....162...14A}
}

@ARTICLE{gbsv3mh,
       author = {{Casamiquela}, L. and {Soubiran}, C. and {Jofr{\'e}}, P. and {Vitali}, S. and {Blanco-Cuaresma}, S. and {Lagarde}, N. and {Slumstrup}, D. and {Heiter}, U. and {Palmerio}, J.~T. and {Brouillet}, N. and {Elgueta}, S. and {Rojas-Arriagada}, A. and {Aguilera-G{\'o}mez}, C. and {Hern{\'a}ndez-Araya}, I. and {Creevey}, O.~L. and {Balaguer-N{\'u}{\~n}ez}, L. and {Carrera}, R.},
        title = "{Gaia FGK benchmark stars: Spectral library and abundances of {\ensuremath{\alpha}} and Fe-peak elements of the third version}",
      journal = {\aap},
         year = 2026,
        month = jan,
       volume = {705},
          eid = {A167},
        pages = {A167},
          doi = {10.1051/0004-6361/202555211},
archivePrefix = {arXiv},
       eprint = {2504.19648},
 primaryClass = {astro-ph.SR},
       adsurl = {https://ui.adsabs.harvard.edu/abs/2026A&A...705A.167C}
}

@ARTICLE{Jofre2014,
       author = {{Jofr{\'e}}, P. and {Heiter}, U. and {Soubiran}, C. and {Blanco-Cuaresma}, S. and {Worley}, C.~C. and {Pancino}, E. and {Cantat-Gaudin}, T. and {Magrini}, L. and {Bergemann}, M. and {Gonz{\'a}lez Hern{\'a}ndez}, J.~I. and {Hill}, V. and {Lardo}, C. and {de Laverny}, P. and {Lind}, K. and {Masseron}, T. and {Montes}, D. and {Mucciarelli}, A. and {Nordlander}, T. and {Recio Blanco}, A. and {Sobeck}, J. and {Sordo}, R. and {Sousa}, S.~G. and {Tabernero}, H. and {Vallenari}, A. and {Van Eck}, S.},
        title = "{Gaia FGK benchmark stars: Metallicity}",
      journal = {\aap},
         year = 2014,
        month = apr,
       volume = {564},
          eid = {A133},
        pages = {A133},
          doi = {10.1051/0004-6361/201322440},
archivePrefix = {arXiv},
       eprint = {1309.1099},
 primaryClass = {astro-ph.GA},
       adsurl = {https://ui.adsabs.harvard.edu/abs/2014A&A...564A.133J}
}

@ARTICLE{basti2018,
       author = {{Hidalgo}, Sebastian L. and {Pietrinferni}, Adriano and {Cassisi}, Santi and {Salaris}, Maurizio and {Mucciarelli}, Alessio and {Savino}, Alessandro and {Aparicio}, Antonio and {Silva Aguirre}, Victor and {Verma}, Kuldeep},
        title = "{The Updated BaSTI Stellar Evolution Models and Isochrones. I. Solar-scaled Calculations}",
      journal = {\apj},
         year = 2018,
        month = apr,
       volume = {856},
       number = {2},
          eid = {125},
        pages = {125},
          doi = {10.3847/1538-4357/aab158},
archivePrefix = {arXiv},
       eprint = {1802.07319},
 primaryClass = {astro-ph.GA},
       adsurl = {https://ui.adsabs.harvard.edu/abs/2018ApJ...856..125H}
}

@ARTICLE{spinsupdate,
       author = {{Casamiquela}, L. and {Reese}, D.~R. and {Lebreton}, Y. and {Haywood}, M. and {Di Matteo}, P. and {Anders}, F. and {Jash}, R. and {Katz}, D. and {Cerqui}, V. and {Boin}, T. and {Kordopatis}, G.},
        title = "{New stellar age estimates using SPInS based on Gaia DR3 photometry and LAMOST DR8 abundances}",
      journal = {\aap},
         year = 2024,
        month = dec,
       volume = {692},
          eid = {A243},
        pages = {A243},
          doi = {10.1051/0004-6361/202451677},
archivePrefix = {arXiv},
       eprint = {2410.15781},
 primaryClass = {astro-ph.GA},
       adsurl = {https://ui.adsabs.harvard.edu/abs/2024A&A...692A.243C}
}

@ARTICLE{flame,
       author = {{Creevey}, O.~L. and {Casamiquela}, L. and {Lebreton}, Y. and
       {Th\'ev\'enin}, F. and {Ordenovic}, C.},
        title = "{Stellar masses and ages in Gaia Data Releases from the Final Luminosity Age Mass Estimator algorithm}",
      journal = {\aap},
         year = 2026,
       volume = {},
          eid = {},
        pages = {},
          doi = {},
archivePrefix = {},
       eprint = {}
}

@software{spinscode,
       author = {{Reese}, Daniel R. and {Lebreton}, Y.},
        title = "{SPInS: Stellar Parameters INferred Systematically}",
 howpublished = {Astrophysics Source Code Library, record ascl:2009.006},
         year = 2020,
        month = sep,
          eid = {ascl:2009.006},
archivePrefix = {ascl},
       eprint = {2009.006},
       adsurl = {https://ui.adsabs.harvard.edu/abs/2020ascl.soft09006R}
}

@ARTICLE{Setterholm2023,
       author = {{Setterholm}, Benjamin R. and {Monnier}, John D. and {Le Bouquin}, Jean-Baptiste and {Anugu}, Narsireddy and {Ennis}, Jacob and {Jocou}, Laurent and {Ibrahim}, Nour and {Kraus}, Stefan and {Anderson}, Matthew D. and {Chhabra}, Sorabh and {Codron}, Isabelle and {Farrington}, Christopher D. and {Flores}, Becky and {Gardner}, Tyler and {Gutierrez}, Mayra and {Lanthermann}, Cyprien and {Majoinen}, Olli W. and {Mortimer}, Daniel J. and {Schaefer}, Gail and {Scott}, Nicholas J. and {ten Brummelaar}, Theo and {Vargas}, Norman L.},
        title = "{MYSTIC: a high angular resolution K-band imager at CHARA}",
      journal = {Journal of Astronomical Telescopes, Instruments, and Systems},
         year = 2023,
        month = apr,
       volume = {9},
          eid = {025006},
        pages = {025006},
          doi = {10.1117/1.JATIS.9.2.025006},
       adsurl = {https://ui.adsabs.harvard.edu/abs/2023JATIS...9b5006S}
}

@ARTICLE{Davis2000,
       author = {{Davis}, J. and {Tango}, W.~J. and {Booth}, A.~J.},
        title = "{Limb-darkening corrections for interferometric uniform disc stellar angular diameters}",
      journal = {\mnras},
         year = 2000,
        month = oct,
       volume = {318},
       number = {2},
        pages = {387-392},
          doi = {10.1046/j.1365-8711.2000.03701.x},
       adsurl = {https://ui.adsabs.harvard.edu/abs/2000MNRAS.318..387D}
}

@ARTICLE{Claret2000,
       author = {{Claret}, A.},
        title = "{A new non-linear limb-darkening law for LTE stellar atmosphere models. Calculations for -5.0 <= log[M/H] <= +1, 2000 K <= T$_{eff}$ <= 50000 K at several surface gravities}",
      journal = {\aap},
         year = 2000,
        month = nov,
       volume = {363},
        pages = {1081-1190},
       adsurl = {https://ui.adsabs.harvard.edu/abs/2000A&A...363.1081C}
}

@ARTICLE{Diaz-Cordoves1992,
       author = {{Diaz-Cordoves}, J. and {Gimenez}, A.},
        title = "{A new nonlinear approximation to the limb-darkening of hot stars}",
      journal = {\aap},
         year = 1992,
        month = jun,
       volume = {259},
       number = {1},
        pages = {227-231},
       adsurl = {https://ui.adsabs.harvard.edu/abs/1992A&A...259..227D}
}

@ARTICLE{heiter2015,
       author = {{Heiter}, U. and {Jofr{\'e}}, P. and {Gustafsson}, B. and {Korn}, A.~J. and {Soubiran}, C. and {Th{\'e}venin}, F.},
        title = "{Gaia FGK benchmark stars: Effective temperatures and surface gravities}",
      journal = {\aap},
         year = 2015,
        month = oct,
       volume = {582},
          eid = {A49},
        pages = {A49},
          doi = {10.1051/0004-6361/201526319},
archivePrefix = {arXiv},
       eprint = {1506.06095},
 primaryClass = {astro-ph.SR},
       adsurl = {https://ui.adsabs.harvard.edu/abs/2015A&A...582A..49H}
}

@ARTICLE{gent2022,
       author = {{Gent}, Matthew Raymond and {Bergemann}, Maria and {Serenelli}, Aldo and {Casagrande}, Luca and {Gerber}, Jeffrey M. and {Heiter}, Ulrike and {Kovalev}, Mikhail and {Morel}, Thierry and {Nardetto}, Nicolas and {Adibekyan}, Vardan and {Silva Aguirre}, V{\'\i}ctor and {Asplund}, Martin and {Belkacem}, Kevin and {del Burgo}, Carlos and {Bigot}, Lionel and {Chiavassa}, Andrea and {Rodr{\'\i}guez D{\'\i}az}, Luisa Fernanda and {Goupil}, Marie-Jo and {Gonz{\'a}lez Hern{\'a}ndez}, Jonay I. and {Mourard}, Denis and {Merle}, Thibault and {M{\'e}sz{\'a}ros}, Szabolcs and {Marshall}, Douglas J. and {Ouazzani}, Rhita-Maria and {Plez}, Bertrand and {Reese}, Daniel and {Trampedach}, Regner and {Tsantaki}, Maria},
        title = "{The SAPP pipeline for the determination of stellar abundances and atmospheric parameters of stars in the core program of the PLATO mission}",
      journal = {\aap},
         year = 2022,
        month = feb,
       volume = {658},
          eid = {A147},
        pages = {A147},
          doi = {10.1051/0004-6361/202140863},
archivePrefix = {arXiv},
       eprint = {2111.06666},
 primaryClass = {astro-ph.SR},
       adsurl = {https://ui.adsabs.harvard.edu/abs/2022A&A...658A.147G}
}

@ARTICLE{Claret2022,
       author = {{Claret}, A. and {Southworth}, J.},
        title = "{Power-2 limb-darkening coefficients for the uvby, UBVRIJHK, SDSS ugriz, Gaia, Kepler, and TESS photometric systems. I. ATLAS stellar atmosphere models}",
      journal = {\aap},
         year = 2022,
        month = aug,
       volume = {664},
          eid = {A128},
        pages = {A128},
          doi = {10.1051/0004-6361/202243827},
archivePrefix = {arXiv},
       eprint = {2206.11098},
 primaryClass = {astro-ph.SR},
       adsurl = {https://ui.adsabs.harvard.edu/abs/2022A&A...664A.128C}
}

@ARTICLE{forman2013PASP..125..306F,
       author = {{Foreman-Mackey}, Daniel and {Hogg}, David W. and {Lang}, Dustin and {Goodman}, Jonathan},
        title = "{emcee: The MCMC Hammer}",
      journal = {\pasp},
         year = 2013,
        month = mar,
       volume = {125},
       number = {925},
        pages = {306},
          doi = {10.1086/670067},
archivePrefix = {arXiv},
       eprint = {1202.3665},
 primaryClass = {astro-ph.IM},
       adsurl = {https://ui.adsabs.harvard.edu/abs/2013PASP..125..306F}
}

@MISC{Plez2012,
       author = {{Plez}, B.},
        title = "{Turbospectrum: Code for spectral synthesis}",
 howpublished = {Astrophysics Source Code Library, record ascl:1205.004},
         year = 2012,
        month = may,
          eid = {ascl:1205.004},
        pages = {ascl:1205.004},
archivePrefix = {ascl},
       eprint = {1205.004},
       adsurl = {https://ui.adsabs.harvard.edu/abs/2012ascl.soft05004P}
}

@ARTICLE{gerber2023,
       author = {{Gerber}, Jeffrey M. and {Magg}, Ekaterina and {Plez}, Bertrand and {Bergemann}, Maria and {Heiter}, Ulrike and {Olander}, Terese and {Hoppe}, Richard},
        title = "{Non-LTE radiative transfer with Turbospectrum}",
      journal = {\aap},
         year = 2023,
        month = jan,
       volume = {669},
          eid = {A43},
        pages = {A43},
          doi = {10.1051/0004-6361/202243673},
archivePrefix = {arXiv},
       eprint = {2206.00967},
 primaryClass = {astro-ph.SR},
       adsurl = {https://ui.adsabs.harvard.edu/abs/2023A&A...669A..43G}
}

@ARTICLE{Espinoza2016,
       author = {{Espinoza}, N{\'e}stor and {Jord{\'a}n}, Andr{\'e}s},
        title = "{Limb darkening and exoplanets - II. Choosing the best law for optimal retrieval of transit parameters}",
      journal = {\mnras},
         year = 2016,
        month = apr,
       volume = {457},
       number = {4},
        pages = {3573-3581},
          doi = {10.1093/mnras/stw224},
archivePrefix = {arXiv},
       eprint = {1601.05485},
 primaryClass = {astro-ph.EP},
       adsurl = {https://ui.adsabs.harvard.edu/abs/2016MNRAS.457.3573E}
}

@ARTICLE{Morello2020,
       author = {{Morello}, G. and {Claret}, A. and {Martin-Lagarde}, M. and {Cossou}, C. and {Tsiaras}, A. and {Lagage}, P. -O.},
        title = "{The ExoTETHyS Package: Tools for Exoplanetary Transits around Host Stars}",
      journal = {\aj},
         year = 2020,
        month = feb,
       volume = {159},
       number = {2},
          eid = {75},
        pages = {75},
          doi = {10.3847/1538-3881/ab63dc},
archivePrefix = {arXiv},
       eprint = {1908.09599},
 primaryClass = {astro-ph.EP},
       adsurl = {https://ui.adsabs.harvard.edu/abs/2020AJ....159...75M}
}

@ARTICLE{Claret2011,
       author = {{Claret}, A. and {Bloemen}, S.},
        title = "{Gravity and limb-darkening coefficients for the Kepler, CoRoT, Spitzer, uvby, UBVRIJHK, and Sloan photometric systems}",
      journal = {\aap},
         year = 2011,
        month = may,
       volume = {529},
          eid = {A75},
        pages = {A75},
          doi = {10.1051/0004-6361/201116451},
       adsurl = {https://ui.adsabs.harvard.edu/abs/2011A&A...529A..75C}
}

@ARTICLE{Maxted2018,
       author = {{Maxted}, P.~F.~L.},
        title = "{Comparison of the power-2 limb-darkening law from the STAGGER-grid to Kepler light curves of transiting exoplanets}",
      journal = {\aap},
         year = 2018,
        month = aug,
       volume = {616},
          eid = {A39},
        pages = {A39},
          doi = {10.1051/0004-6361/201832944},
archivePrefix = {arXiv},
       eprint = {1804.07943},
 primaryClass = {astro-ph.SR},
       adsurl = {https://ui.adsabs.harvard.edu/abs/2018A&A...616A..39M}
}

@ARTICLE{Anugu2020,
       author = {{Anugu}, Narsireddy and {Le Bouquin}, Jean-Baptiste and {Monnier}, John D. and {Kraus}, Stefan and {Setterholm}, Benjamin R. and {Labdon}, Aaron and {Davies}, Claire L. and {Lanthermann}, Cyprien and {Gardner}, Tyler and {Ennis}, Jacob and {Johnson}, Keith J.~C. and {Ten Brummelaar}, Theo and {Schaefer}, Gail and {Sturmann}, Judit},
        title = "{MIRC-X: A Highly Sensitive Six-telescope Interferometric Imager at the CHARA Array}",
      journal = {\aj},
         year = 2020,
        month = oct,
       volume = {160},
       number = {4},
          eid = {158},
        pages = {158},
          doi = {10.3847/1538-3881/aba957},
archivePrefix = {arXiv},
       eprint = {2007.12320},
 primaryClass = {astro-ph.IM},
       adsurl = {https://ui.adsabs.harvard.edu/abs/2020AJ....160..158A}
}

@ARTICLE{Ebrahimkutty2024,
       author = {{Ebrahimkutty}, N. and {Gent}, M.~R. and {Mourard}, D. and {Domiciano de Souza}, A. and {Bergemann}, M. and {Morel}, T. and {Morello}, G. and {Nardetto}, N. and {Plez}, B.},
        title = "{Optimised use of interferometry, spectroscopy, and stellar atmosphere models for determining the fundamental parameters of stars}",
      journal = {\aap},
         year = 2024,
        month = nov,
       volume = {691},
          eid = {A207},
        pages = {A207},
          doi = {10.1051/0004-6361/202450105},
       adsurl = {https://ui.adsabs.harvard.edu/abs/2024A&A...691A.207E}
}

@INPROCEEDINGS{Dekkar2000SPIE.4008..534D,
       author = {{Dekker}, Hans and {D'Odorico}, Sandro and {Kaufer}, Andreas and {Delabre}, Bernard and {Kotzlowski}, Heinz},
        title = "{Design, construction, and performance of UVES, the echelle spectrograph for the UT2 Kueyen Telescope at the ESO Paranal Observatory}",
    booktitle = {Optical and IR Telescope Instrumentation and Detectors},
         year = 2000,
       editor = {{Iye}, Masanori and {Moorwood}, Alan F.},
       series = {Society of Photo-Optical Instrumentation Engineers (SPIE) Conference Series},
       volume = {4008},
        month = aug,
        pages = {534-545},
          doi = {10.1117/12.395512},
       adsurl = {https://ui.adsabs.harvard.edu/abs/2000SPIE.4008..534D}
}

@ARTICLE{mayor2003Msngr.114...20M,
       author = {{Mayor}, M. and {Pepe}, F. and {Queloz}, D. and {Bouchy}, F. and {Rupprecht}, G. and {Lo Curto}, G. and {Avila}, G. and {Benz}, W. and {Bertaux}, J. -L. and {Bonfils}, X. and {Dall}, Th. and {Dekker}, H. and {Delabre}, B. and {Eckert}, W. and {Fleury}, M. and {Gilliotte}, A. and {Gojak}, D. and {Guzman}, J.~C. and {Kohler}, D. and {Lizon}, J. -L. and {Longinotti}, A. and {Lovis}, C. and {Megevand}, D. and {Pasquini}, L. and {Reyes}, J. and {Sivan}, J. -P. and {Sosnowska}, D. and {Soto}, R. and {Udry}, S. and {van Kesteren}, A. and {Weber}, L. and {Weilenmann}, U.},
        title = "{Setting New Standards with HARPS}",
      journal = {The Messenger},
         year = 2003,
        month = dec,
       volume = {114},
        pages = {20-24},
       adsurl = {https://ui.adsabs.harvard.edu/abs/2003Msngr.114...20M}
}

@INPROCEEDINGS{Haubois2014SPIE.9146E..0OH,
       author = {{Haubois}, Xavier and {Bernaud}, Patrick and {Mella}, Guillaume and {Duvert}, Gilles and {Benisty}, Myriam and {B{\'e}rio}, Philippe and {Bourges}, Laurent and {Chelli}, Alain E. and {Chesneau}, Olivier and {Lacour}, Sylvestre and {Lafrasse}, Sylvain and {Le Bouquin}, Jean-Baptiste and {Mourard}, Denis and {Nardetto}, Nicolas and {Olofsson}, Johan},
        title = "{A global database for optical interferometry}",
    booktitle = {Optical and Infrared Interferometry IV},
         year = 2014,
       editor = {{Rajagopal}, Jayadev K. and {Creech-Eakman}, Michelle J. and {Malbet}, Fabien},
       series = {Society of Photo-Optical Instrumentation Engineers (SPIE) Conference Series},
       volume = {9146},
        month = jul,
          eid = {91460O},
        pages = {91460O},
          doi = {10.1117/12.2056977},
       adsurl = {https://ui.adsabs.harvard.edu/abs/2014SPIE.9146E..0OH}
}

@ARTICLE{scipy,
  author  = {Virtanen, Pauli and Gommers, Ralf and Oliphant, Travis E. and
            Haberland, Matt and Reddy, Tyler and Cournapeau, David and
            Burovski, Evgeni and Peterson, Pearu and Weckesser, Warren and
            Bright, Jonathan and {van der Walt}, St{\'e}fan J. and
            Brett, Matthew and Wilson, Joshua and Millman, K. Jarrod and
            Mayorov, Nikolay and Nelson, Andrew R. J. and Jones, Eric and
            Kern, Robert and Larson, Eric and Carey, C J and
            Polat, {\.I}lhan and Feng, Yu and Moore, Eric W. and
            {VanderPlas}, Jake and Laxalde, Denis and Perktold, Josef and
            Cimrman, Robert and Henriksen, Ian and Quintero, E. A. and
            Harris, Charles R. and Archibald, Anne M. and
            Ribeiro, Ant{\^o}nio H. and Pedregosa, Fabian and
            {van Mulbregt}, Paul and {SciPy 1.0 Contributors}},
  title   = {{{SciPy} 1.0: Fundamental Algorithms for Scientific
            Computing in Python}},
  journal = {Nature Methods},
  year    = {2020},
  volume  = {17},
  pages   = {261--272},
  adsurl  = {https://rdcu.be/b08Wh},
  doi     = {10.1038/s41592-019-0686-2},
}

@ARTICLE{gaiadr3,
       author = {{Gaia Collaboration} and {Vallenari}, A. and {Brown}, A.~G.~A. and {Prusti}, T. and {de Bruijne}, J.~H.~J. and {Arenou}, F. and {Babusiaux}, C. and {Biermann}, M. and {Creevey}, O.~L. and {Ducourant}, C. and {Evans}, D.~W. and {Eyer}, L. and {Guerra}, R. and {Hutton}, A. and {Jordi}, C. and {Klioner}, S.~A. and {Lammers}, U.~L. and {Lindegren}, L. and {Luri}, X. and {Mignard}, F. and {Panem}, C. and {Pourbaix}, D. and {Randich}, S. and {Sartoretti}, P. and {Soubiran}, C. and {Tanga}, P. and {Walton}, N.~A. and {Bailer-Jones}, C.~A.~L. and {Bastian}, U. and {Drimmel}, R. and {Jansen}, F. and {Katz}, D. and {Lattanzi}, M.~G. and {van Leeuwen}, F. and {Bakker}, J. and {Cacciari}, C. and {Casta{\~n}eda}, J. and {De Angeli}, F. and {Fabricius}, C. and {Fouesneau}, M. and {Fr{\'e}mat}, Y. and {Galluccio}, L. and {Guerrier}, A. and {Heiter}, U. and {Masana}, E. and {Messineo}, R. and {Mowlavi}, N. and {Nicolas}, C. and {Nienartowicz}, K. and {Pailler}, F. and {Panuzzo}, P. and {Riclet}, F. and {Roux}, W. and {Seabroke}, G.~M. and {Sordo}, R. and {Th{\'e}venin}, F. and {Gracia-Abril}, G. and {Portell}, J. and {Teyssier}, D. and {Altmann}, M. and {Andrae}, R. and {Audard}, M. and {Bellas-Velidis}, I. and {Benson}, K. and {Berthier}, J. and {Blomme}, R. and {Burgess}, P.~W. and {Busonero}, D. and {Busso}, G. and {C{\'a}novas}, H. and {Carry}, B. and {Cellino}, A. and {Cheek}, N. and {Clementini}, G. and {Damerdji}, Y. and {Davidson}, M. and {de Teodoro}, P. and {Nu{\~n}ez Campos}, M. and {Delchambre}, L. and {Dell'Oro}, A. and {Esquej}, P. and {Fern{\'a}ndez-Hern{\'a}ndez}, J. and {Fraile}, E. and {Garabato}, D. and {Garc{\'\i}a-Lario}, P. and {Gosset}, E. and {Haigron}, R. and {Halbwachs}, J.-L. and {Hambly}, N.~C. and {Harrison}, D.~L. and {Hern{\'a}ndez}, J. and {Hestroffer}, D. and {Hodgkin}, S.~T. and {Holl}, B. and {Jan{\ss}en}, K. and {Jevardat de Fombelle}, G. and {Jordan}, S. and {Krone-Martins}, A. and {Lanzafame}, A.~C. and {L{\"o}ffler}, W. and {Marchal}, O. and {Marrese}, P.~M. and {Moitinho}, A. and {Muinonen}, K. and {Osborne}, P. and {Pancino}, E. and {Pauwels}, T. and {Recio-Blanco}, A. and {Reyl{\'e}}, C. and {Riello}, M. and {Rimoldini}, L. and {Roegiers}, T. and {Rybizki}, J. and {Sarro}, L.~M. and {Siopis}, C. and {Smith}, M. and {Sozzetti}, A. and {Utrilla}, E. and {van Leeuwen}, M. and {Abbas}, U. and {{\'A}brah{\'a}m}, P. and {Abreu Aramburu}, A. and {Aerts}, C. and {Aguado}, J.~J. and {Ajaj}, M. and {Aldea-Montero}, F. and {Altavilla}, G. and {{\'A}lvarez}, M.~A. and {Alves}, J. and {Anders}, F. and {Anderson}, R.~I. and {Anglada Varela}, E. and {Antoja}, T. and {Baines}, D. and {Baker}, S.~G. and {Balaguer-N{\'u}{\~n}ez}, L. and {Balbinot}, E. and {Balog}, Z. and {Barache}, C. and {Barbato}, D. and {Barros}, M. and {Barstow}, M.~A. and {Bartolom{\'e}}, S. and {Bassilana}, J.-L. and {Bauchet}, N. and {Becciani}, U. and {Bellazzini}, M. and {Berihuete}, A. and {Bernet}, M. and {Bertone}, S. and {Bianchi}, L. and {Binnenfeld}, A. and {Blanco-Cuaresma}, S. and {Blazere}, A. and {Boch}, T. and {Bombrun}, A. and {Bossini}, D. and {Bouquillon}, S. and {Bragaglia}, A. and {Bramante}, L. and {Breedt}, E. and {Bressan}, A. and {Brouillet}, N. and {Brugaletta}, E. and {Bucciarelli}, B. and {Burlacu}, A. and {Butkevich}, A.~G. and {Buzzi}, R. and {Caffau}, E. and {Cancelliere}, R. and {Cantat-Gaudin}, T. and {Carballo}, R. and {Carlucci}, T. and {Carnerero}, M.~I. and {Carrasco}, J.~M. and {Casamiquela}, L. and {Castellani}, M. and {Castro-Ginard}, A. and {Chaoul}, L. and {Charlot}, P. and {Chemin}, L. and {Chiaramida}, V. and {Chiavassa}, A. and {Chornay}, N. and {Comoretto}, G. and {Contursi}, G. and {Cooper}, W.~J. and {Cornez}, T. and {Cowell}, S. and {Crifo}, F. and {Cropper}, M. and {Crosta}, M. and {Crowley}, C. and {Dafonte}, C. and {Dapergolas}, A. and {David}, M. and {David}, P. and {de Laverny}, P. and {De Luise}, F. and {De March}, R.},
        title = "{Gaia Data Release 3. Summary of the content and survey properties}",
      journal = {\aap},
         year = 2023,
        month = jun,
       volume = {674},
          eid = {A1},
        pages = {A1},
          doi = {10.1051/0004-6361/202243940},
archivePrefix = {arXiv},
       eprint = {2208.00211},
 primaryClass = {astro-ph.GA},
       adsurl = {https://ui.adsabs.harvard.edu/abs/2023A&A...674A...1G}
}

@dataset{ducati,
       author = {{Ducati}, J.~R.},
        title = "{VizieR Online Data Catalog: Stellar Photometry in Johnson's 11-color system (Ducati, 2002)}",
 howpublished = {VizieR On-line Data Catalog: II/237.  Originally published in: Departement of Astronomy, University of Wisconsin, Madison WI 53706 (2002)},
         year = 2002,
        month = aug,
          eid = {II/237},
       adsurl = {https://ui.adsabs.harvard.edu/abs/2002yCat.2237....0D}
}

@dataset{2mass,
       author = {{Cutri}, R.~M. and {Skrutskie}, M.~F. and {van Dyk}, S. and {Beichman}, C.~A. and {Carpenter}, J.~M. and {Chester}, T. and {Cambresy}, L. and {Evans}, T. and {Fowler}, J. and {Gizis}, J. and {Howard}, E. and {Huchra}, J. and {Jarrett}, T. and {Kopan}, E.~L. and {Kirkpatrick}, J.~D. and {Light}, R.~M. and {Marsh}, K.~A. and {McCallon}, H. and {Schneider}, S. and {Stiening}, R. and {Sykes}, M. and {Weinberg}, M. and {Wheaton}, W.~A. and {Wheelock}, S. and {Zacarias}, N.},
        title = "{VizieR Online Data Catalog: 2MASS All-Sky Catalog of Point Sources (Cutri+ 2003)}",
 howpublished = {VizieR On-line Data Catalog: II/246.  Originally published in: University of Massachusetts and Infrared Processing and Analysis Center, (IPAC/California Institute of Technology) (2003)},
         year = 2003,
        month = jun,
          eid = {II/246},
       adsurl = {https://ui.adsabs.harvard.edu/abs/2003yCat.2246....0C}
}

@ARTICLE{Oja1993,
       author = {{Oja}, T.},
        title = "{UBV photometry of stars whose positions are accurately known. VII.}",
      journal = {\aaps},
         year = 1993,
        month = sep,
       volume = {100},
        pages = {591-592},
       adsurl = {https://ui.adsabs.harvard.edu/abs/1993A&AS..100..591O}
}

@ARTICLE{vanbelle2009,
       author = {{van Belle}, Gerard T. and {von Braun}, Kaspar},
        title = "{Directly Determined Linear Radii and Effective Temperatures of Exoplanet Host Stars}",
      journal = {\apj},
         year = 2009,
        month = apr,
       volume = {694},
       number = {2},
        pages = {1085-1098},
          doi = {10.1088/0004-637X/694/2/1085},
archivePrefix = {arXiv},
       eprint = {0901.1206},
 primaryClass = {astro-ph.SR},
       adsurl = {https://ui.adsabs.harvard.edu/abs/2009ApJ...694.1085V}
}

@ARTICLE{Laney2012,
       author = {{Laney}, C.~D. and {Joner}, M.~D. and {Pietrzy{\'n}ski}, G.},
        title = "{A new Large Magellanic Cloud K-band distance from precision measurements of nearby red clump stars}",
      journal = {\mnras},
         year = 2012,
        month = jan,
       volume = {419},
       number = {2},
        pages = {1637-1641},
          doi = {10.1111/j.1365-2966.2011.19826.x},
archivePrefix = {arXiv},
       eprint = {1109.4800},
 primaryClass = {astro-ph.SR},
       adsurl = {https://ui.adsabs.harvard.edu/abs/2012MNRAS.419.1637L}
}

@ARTICLE{Harps2003,
       author = {{Mayor}, M. and {Pepe}, F. and {Queloz}, D. and {Bouchy}, F. and {Rupprecht}, G. and {Lo Curto}, G. and {Avila}, G. and {Benz}, W. and {Bertaux}, J.-L. and {Bonfils}, X. and {Dall}, Th. and {Dekker}, H. and {Delabre}, B. and {Eckert}, W. and {Fleury}, M. and {Gilliotte}, A. and {Gojak}, D. and {Guzman}, J.~C. and {Kohler}, D. and {Lizon}, J.-L. and {Longinotti}, A. and {Lovis}, C. and {Megevand}, D. and {Pasquini}, L. and {Reyes}, J. and {Sivan}, J.-P. and {Sosnowska}, D. and {Soto}, R. and {Udry}, S. and {van Kesteren}, A. and {Weber}, L. and {Weilenmann}, U.},
        title = "{Setting New Standards with HARPS}",
      journal = {The Messenger},
         year = 2003,
        month = dec,
       volume = {114},
        pages = {20-24},
       adsurl = {https://ui.adsabs.harvard.edu/abs/2003Msngr.114...20M}
}

@INPROCEEDINGS{Dekker2000,
       author = {{Dekker}, Hans and {D'Odorico}, Sandro and {Kaufer}, Andreas and {Delabre}, Bernard and {Kotzlowski}, Heinz},
        title = "{Design, construction, and performance of UVES, the echelle spectrograph for the UT2 Kueyen Telescope at the ESO Paranal Observatory}",
    booktitle = {Optical and IR Telescope Instrumentation and Detectors},
         year = 2000,
       editor = {{Iye}, Masanori and {Moorwood}, Alan F.},
       series = {Society of Photo-Optical Instrumentation Engineers (SPIE) Conference Series},
       volume = {4008},
        month = aug,
        pages = {534-545},
          doi = {10.1117/12.395512},
       adsurl = {https://ui.adsabs.harvard.edu/abs/2000SPIE.4008..534D}
}

@INPROCEEDINGS{Bourg2016,
       author = {{Bourg{\`e}s}, L. and {Duvert}, G.},
        title = "{ASPRO2: get ready for VLTI's instruments GRAVITY and MATISSE}",
    booktitle = {Optical and Infrared Interferometry and Imaging V},
         year = 2016,
       editor = {{Malbet}, Fabien and {Creech-Eakman}, Michelle J. and {Tuthill}, Peter G.},
       series = {Society of Photo-Optical Instrumentation Engineers (SPIE) Conference Series},
       volume = {9907},
        month = aug,
          eid = {990711},
        pages = {990711},
          doi = {10.1117/12.2234426},
       adsurl = {https://ui.adsabs.harvard.edu/abs/2016SPIE.9907E..11B}
}

@ARTICLE{Perrin2003,
       author = {{Perrin}, G.},
        title = "{The calibration of interferometric visibilities obtained  with single-mode optical interferometers. Computation of error bars and correlations}",
      journal = {\aap},
         year = 2003,
        month = mar,
       volume = {400},
        pages = {1173-1181},
          doi = {10.1051/0004-6361:20030025},
archivePrefix = {arXiv},
       eprint = {astro-ph/0301140},
 primaryClass = {astro-ph},
       adsurl = {https://ui.adsabs.harvard.edu/abs/2003A&A...400.1173P}
}

@ARTICLE{Lachaume2019,
       author = {{Lachaume}, R{\'e}gis and {Rabus}, Markus and {Jord{\'a}n}, Andr{\'e}s and {Brahm}, Rafael and {Boyajian}, Tabetha and {von Braun}, Kaspar and {Berger}, Jean-Philippe},
        title = "{Towards reliable uncertainties in IR interferometry: the bootstrap for correlated statistical and systematic errors}",
      journal = {\mnras},
         year = 2019,
        month = apr,
       volume = {484},
       number = {2},
        pages = {2656-2673},
          doi = {10.1093/mnras/stz114},
archivePrefix = {arXiv},
       eprint = {1901.02879},
 primaryClass = {astro-ph.IM},
       adsurl = {https://ui.adsabs.harvard.edu/abs/2019MNRAS.484.2656L}
}

@ARTICLE{Paegert2021,
       author = {{Paegert}, Martin and {Stassun}, Keivan G. and {Collins}, Karen A. and {Pepper}, Joshua and {Torres}, Guillermo and {Jenkins}, Jon and {Twicken}, Joseph D. and {Latham}, David W.},
        title = "{TESS Input Catalog versions 8.1 and 8.2: Phantoms in the 8.0 Catalog and How to Handle Them}",
      journal = {arXiv e-prints},
         year = 2021,
        month = aug,
          eid = {arXiv:2108.04778},
        pages = {arXiv:2108.04778},
          doi = {10.48550/arXiv.2108.04778},
archivePrefix = {arXiv},
       eprint = {2108.04778},
 primaryClass = {astro-ph.EP},
       adsurl = {https://ui.adsabs.harvard.edu/abs/2021arXiv210804778P}
}

@ARTICLE{Allende1999,
       author = {{Allende Prieto}, C. and {Lambert}, D.~L.},
        title = "{Fundamental parameters of nearby stars from the comparison with evolutionary calculations: masses, radii and effective temperatures}",
      journal = {\aap},
         year = 1999,
        month = dec,
       volume = {352},
        pages = {555-562},
          doi = {10.48550/arXiv.astro-ph/9911002},
archivePrefix = {arXiv},
       eprint = {astro-ph/9911002},
 primaryClass = {astro-ph},
       adsurl = {https://ui.adsabs.harvard.edu/abs/1999A&A...352..555A}
}

@ARTICLE{Hekker2006,
       author = {{Hekker}, S. and {Reffert}, S. and {Quirrenbach}, A. and {Mitchell}, D.~S. and {Fischer}, D.~A. and {Marcy}, G.~W. and {Butler}, R.~P.},
        title = "{Precise radial velocities of giant stars. I. Stable stars}",
      journal = {\aap},
         year = 2006,
        month = aug,
       volume = {454},
       number = {3},
        pages = {943-949},
          doi = {10.1051/0004-6361:20064946},
archivePrefix = {arXiv},
       eprint = {astro-ph/0604502},
 primaryClass = {astro-ph},
       adsurl = {https://ui.adsabs.harvard.edu/abs/2006A&A...454..943H}
}

@ARTICLE{simbad,
       author = {{Wenger}, M. and {Ochsenbein}, F. and {Egret}, D. and {Dubois}, P. and {Bonnarel}, F. and {Borde}, S. and {Genova}, F. and {Jasniewicz}, G. and {Lalo{\"e}}, S. and {Lesteven}, S. and {Monier}, R.},
        title = "{The SIMBAD astronomical database. The CDS reference database for astronomical objects}",
      journal = {\aaps},
         year = 2000,
        month = apr,
       volume = {143},
        pages = {9-22},
          doi = {10.1051/aas:2000332},
archivePrefix = {arXiv},
       eprint = {astro-ph/0002110},
 primaryClass = {astro-ph},
       adsurl = {https://ui.adsabs.harvard.edu/abs/2000A&AS..143....9W}
}

@ARTICLE{Teunissen2008,
       author = {{Teunissen}, P.~J.~G. and {Amiri-Simkooei}, A.~R.},
        title = "{Least-squares variance component estimation}",
      journal = {Journal of Geodesy},
         year = 2008,
        month = feb,
       volume = {82},
       number = {2},
        pages = {65-82},
          doi = {10.1007/s00190-007-0157-x},
       adsurl = {https://ui.adsabs.harvard.edu/abs/2008JGeod..82...65T}
}

@ARTICLE{Creevey2023,
       author = {{Creevey}, O.~L. and {Sordo}, R. and {Pailler}, F. and {Fr{\'e}mat}, Y. and {Heiter}, U. and {Th{\'e}venin}, F. and {Andrae}, R. and {Fouesneau}, M. and {Lobel}, A. and {Bailer-Jones}, C.~A.~L. and {Garabato}, D. and {Bellas-Velidis}, I. and {Brugaletta}, E. and {Lorca}, A. and {Ordenovic}, C. and {Palicio}, P.~A. and {Sarro}, L.~M. and {Delchambre}, L. and {Drimmel}, R. and {Rybizki}, J. and {Torralba Elipe}, G. and {Korn}, A.~J. and {Recio-Blanco}, A. and {Schultheis}, M.~S. and {De Angeli}, F. and {Montegriffo}, P. and {Abreu Aramburu}, A. and {Accart}, S. and {{\'A}lvarez}, M.~A. and {Bakker}, J. and {Brouillet}, N. and {Burlacu}, A. and {Carballo}, R. and {Casamiquela}, L. and {Chiavassa}, A. and {Contursi}, G. and {Cooper}, W.~J. and {Dafonte}, C. and {Dapergolas}, A. and {de Laverny}, P. and {Dharmawardena}, T.~E. and {Edvardsson}, B. and {Le Fustec}, Y. and {Garc{\'\i}a-Lario}, P. and {Garc{\'\i}a-Torres}, M. and {Gomez}, A. and {Gonz{\'a}lez-Santamar{\'\i}a}, I. and {Hatzidimitriou}, D. and {Jean-Antoine Piccolo}, A. and {Kontiza}, M. and {Kordopatis}, G. and {Lanzafame}, A.~C. and {Lebreton}, Y. and {Licata}, E.~L. and {Lindstr{\o}m}, H.~E.~P. and {Livanou}, E. and {Magdaleno Romeo}, A. and {Manteiga}, M. and {Marocco}, F. and {Marshall}, D.~J. and {Mary}, N. and {Nicolas}, C. and {Pallas-Quintela}, L. and {Panem}, C. and {Pichon}, B. and {Poggio}, E. and {Riclet}, F. and {Robin}, C. and {Santove{\~n}a}, R. and {Silvelo}, A. and {Slezak}, I. and {Smart}, R.~L. and {Soubiran}, C. and {S{\"u}veges}, M. and {Ulla}, A. and {Utrilla}, E. and {Vallenari}, A. and {Zhao}, H. and {Zorec}, J. and {Barrado}, D. and {Bijaoui}, A. and {Bouret}, J.-C. and {Blomme}, R. and {Brott}, I. and {Cassisi}, S. and {Kochukhov}, O. and {Martayan}, C. and {Shulyak}, D. and {Silvester}, J.},
        title = "{Gaia Data Release 3. Astrophysical parameters inference system (Apsis). I. Methods and content overview}",
      journal = {\aap},
         year = 2023,
        month = jun,
       volume = {674},
          eid = {A26},
        pages = {A26},
          doi = {10.1051/0004-6361/202243688},
archivePrefix = {arXiv},
       eprint = {2206.05864},
 primaryClass = {astro-ph.GA},
       adsurl = {https://ui.adsabs.harvard.edu/abs/2023A&A...674A..26C}
}

@ARTICLE{Mourard2026,
       author = {{Mourard}, D. and {B{\'e}rio}, P. and {Bailet}, C. and {Caci}, A. and {Dejonghe}, J. and {Geneslay}, P. and {Lagarde}, S. and {Lecron}, D. and {Meilland}, A. and {Morand}, F. and {Nardetto}, N. and {Pannetier}, C. and {Perraut}, K. and {Rousseau}, S. and {Salabert}, D. and {Ebrahimkutty}, N. and {Iba{\~n}ez Bustos}, R.~V. and {Jon{\'a}k}, J. and {Ligi}, R. and {Nowacki}, H. and {Patru}, F. and {Vrard}, M. and {Bourg{\'e}s}, L. and {Mella}, G. and {Anugu}, N. and {ten Brummelaar}, T. and {Farrington}, C. and {Gies}, D. and {Jones}, J. and {Koehler}, R. and {Kubiak}, K. and {Lanthermann}, C. and {Ligon}, R. and {Schaefer}, G. and {Scott}, N. and {Turner}, N. and {Monnier}, J. and {Kraus}, S. and {Le Bouquin}, J.~B. and {Gardner}, T. and {Gutierrez}, M. and {Ibrahim}, N. and {Aristidi}, E. and {Caujolle}, Y. and {Giordano}, C. and {Ziad}, A.},
        title = "{CHARA/SPICA: The six-telescope visible combiner and near-infrared fringe tracker for the CHARA Array}",
      journal = {arXiv e-prints},
         year = 2026,
        month = sep,
          eid = {arXiv:2609.02406},
        pages = {arXiv:2609.02406},
          doi = {10.48550/arXiv.2609.02406},
archivePrefix = {arXiv},
       eprint = {2609.02406},
 primaryClass = {astro-ph.IM},
       adsurl = {https://ui.adsabs.harvard.edu/abs/2026arXiv260902406M}
}

@ARTICLE{Rauer2025,
       author = {{Rauer}, Heike and {Aerts}, Conny and {Cabrera}, Juan and {Deleuil}, Magali and {Erikson}, Anders and {Gizon}, Laurent and {Goupil}, Mariejo and {Heras}, Ana and {Walloschek}, Thomas and {Lorenzo-Alvarez}, Jose and {Marliani}, Filippo and {Martin-Garcia}, C{\'e}sar and {Mas-Hesse}, J. Miguel and {O'Rourke}, Laurence and {Osborn}, Hugh and {Pagano}, Isabella and {Piotto}, Giampaolo and {Pollacco}, Don and {Ragazzoni}, Roberto and {Ramsay}, Gavin and {Udry}, St{\'e}phane and {Appourchaux}, Thierry and {Benz}, Willy and {Brandeker}, Alexis and {G{\"u}del}, Manuel and {Janot-Pacheco}, Eduardo and {Kabath}, Petr and {Kjeldsen}, Hans and {Min}, Michiel and {Santos}, Nuno and {Smith}, Alan and {Suarez}, Juan-Carlos and {Werner}, Stephanie C. and {Aboudan}, Alessio and {Abreu}, Manuel and {Acu{\~n}a}, Lorena and {Adams}, Moritz and {Adibekyan}, Vardan and {Affer}, Laura and {Agneray}, Fran{\c{c}}ois and {Agnor}, Craig and {Aguirre B{\o}rsen-Koch}, Victor and {Ahmed}, Saad and {Aigrain}, Suzanne and {Al-Bahlawan}, Ashraf and {Alcacera Gil}, Ma de los Angeles and {Alei}, Eleonora and {Alencar}, Silvia and {Alexander}, Richard and {Alfonso-Garz{\'o}n}, Julia and {Alibert}, Yann and {Allende Prieto}, Carlos and {Almeida}, Leonardo and {Alonso Sobrino}, Roi and {Altavilla}, Giuseppe and {Althaus}, Christian and {Alvarez Trujillo}, Luis Alonso and {Amarsi}, Anish and {Ammler-von Eiff}, Matthias and {Am{\^o}res}, Eduardo and {Andrade}, Laerte and {Antoniadis-Karnavas}, Alexandros and {Ant{\'o}nio}, Carlos and {Aparicio del Moral}, Beatriz and {Appolloni}, Matteo and {Arena}, Claudio and {Armstrong}, David and {Aroca Aliaga}, Jose and {Asplund}, Martin and {Audenaert}, Jeroen and {Auricchio}, Natalia and {Avelino}, Pedro and {Baeke}, Ann and {Bailli{\'e}}, Kevin and {Balado}, Ana and {Ballber Balaguer{\'o}}, Pau and {Balestra}, Andrea and {Ball}, Warrick and {Ballans}, Herve and {Ballot}, Jerome and {Barban}, Caroline and {Barbary}, Ga{\"e}le and {Barbieri}, Mauro and {Barcel{\'o} Forteza}, Sebasti{\`a} and {Barker}, Adrian and {Barklem}, Paul and {Barnes}, Sydney and {Barrado Navascues}, David and {Barragan}, Oscar and {Baruteau}, Cl{\'e}ment and {Basu}, Sarbani and {Baudin}, Frederic and {Baumeister}, Philipp and {Bayliss}, Daniel and {Bazot}, Michael and {Beck}, Paul G. and {Belkacem}, Kevin and {Bellinger}, Earl and {Benatti}, Serena and {Benomar}, Othman and {B{\'e}rard}, Diane and {Bergemann}, Maria and {Bergomi}, Maria and {Bernardo}, Pierre and {Biazzo}, Katia and {Bignamini}, Andrea and {Bigot}, Lionel and {Billot}, Nicolas and {Binet}, Martin and {Biondi}, David and {Biondi}, Federico and {Birch}, Aaron C. and {Bitsch}, Bertram and {Bluhm Ceballos}, Paz Victoria and {B{\'o}di}, Attila and {Bogn{\'a}r}, Zs{\'o}fia and {Boisse}, Isabelle and {Bolmont}, Emeline and {Bonanno}, Alfio and {Bonavita}, Mariangela and {Bonfanti}, Andrea and {Bonfils}, Xavier and {Bonito}, Rosaria and {Bonomo}, Aldo Stefano and {B{\"o}rner}, Anko and {Boro Saikia}, Sudeshna and {Borreguero Mart{\'\i}n}, Elisa and {Borsa}, Francesco and {Borsato}, Luca and {Bossini}, Diego and {Bouchy}, Francois and {Bou{\'e}}, Gwena{\"e}l and {Boufleur}, Rodrigo and {Boumier}, Patrick and {Bourrier}, Vincent and {Bowman}, Dominic M. and {Bozzo}, Enrico and {Bradley}, Louisa and {Bray}, John and {Bressan}, Alessandro and {Breton}, Sylvain and {Brienza}, Daniele and {Brito}, Ana and {Brogi}, Matteo and {Brown}, Beverly and {Brown}, David J.~A. and {Brun}, Allan Sacha and {Bruno}, Giovanni and {Bruns}, Michael and {Buchhave}, Lars A. and {Bugnet}, Lisa and {Buldgen}, Ga{\"e}l and {Burgess}, Patrick and {Busatta}, Andrea and {Busso}, Giorgia and {Buzasi}, Derek and {Caballero}, Jos{\'e} A. and {Cabral}, Alexandre and {Cabrero Gomez}, Juan-Francisco and {Calderone}, Flavia and {Cameron}, Robert and {Cameron}, Andrew and {Campante}, Tiago and {Campos Gestal}, N{\'e}stor and {Canto Martins}, Bruno Leonardo and {Cara}, Christophe and {Carone}, Ludmila and {Carrasco}, Josep Manel and {Casagrande}, Luca and {Casewell}, Sarah L. and {Cassisi}, Santi and {Castellani}, Marco and {Castro}, Matthieu and {Catala}, Claude and {Catal{\'a}n Fern{\'a}ndez}, Irene and {Catelan}, M{\'a}rcio and {Cegla}, Heather and {Cerruti}, Chiara and {Cessa}, Virginie and {Chadid}, Merieme and {Chaplin}, William and {Charpinet}, Stephane and {Chiappini}, Cristina and {Chiarucci}, Simone and {Chiavassa}, Andrea and {Chinellato}, Simonetta and {Chirulli}, Giovanni and {Christensen-Dalsgaard}, J{\o}rgen and {Church}, Ross and {Claret}, Antonio and {Clarke}, Cathie and {Claudi}, Riccardo and {Clermont}, Lionel and {Coelho}, Hugo and {Coelho}, Joao and {Cogato}, Fabrizio and {Colom{\'e}}, Josep and {Condamin}, Mathieu and {Conde Garc{\'\i}a}, Fernando and {Conseil}, Simon},
        title = "{The PLATO mission}",
      journal = {Experimental Astronomy},
         year = 2025,
        month = jun,
       volume = {59},
       number = {3},
          eid = {26},
        pages = {26},
          doi = {10.1007/s10686-025-09985-9},
archivePrefix = {arXiv},
       eprint = {2406.05447},
 primaryClass = {astro-ph.IM},
       adsurl = {https://ui.adsabs.harvard.edu/abs/2025ExA....59...26R}
}

@ARTICLE{Montalto2026,
       author = {{Montalto}, M. and {Piotto}, G. and {Marrese}, P.~M. and {Prisinzano}, L. and {Marinoni}, S. and {Granata}, V. and {Cabrera}, J. and {Nascimbeni}, V. and {Desidera}, S. and {Adibekyan}, V. and {Ortolani}, S. and {Alei}, E. and {Aerts}, C. and {Altavilla}, G. and {Belkacem}, K. and {Benatti}, S. and {B{\"o}rner}, A. and {Deleuil}, M. and {Fabrizio}, M. and {Gizon}, L. and {Goupil}, M.~J. and {G{\"u}nther}, M. and {Heras}, A.~M. and {Magrin}, D. and {Malavolta}, L. and {Mas-Hesse}, J.~M. and {Pagano}, I. and {Paproth}, C. and {Pollacco}, D. and {Ragazzoni}, R. and {Ramsay}, G. and {Rauer}, H. and {Udry}, S.},
        title = "{The PLATO Input Catalogue of targets (tPIC) for the first Long Pointing Field}",
      journal = {arXiv e-prints},
         year = 2026,
        month = apr,
          eid = {arXiv:2604.03369},
        pages = {arXiv:2604.03369},
          doi = {10.48550/arXiv.2604.03369},
archivePrefix = {arXiv},
       eprint = {2604.03369},
 primaryClass = {astro-ph.EP},
       adsurl = {https://ui.adsabs.harvard.edu/abs/2026arXiv260403369M}
}

@ARTICLE{Blanco-Cuaresma2019,
       author = {{Blanco-Cuaresma}, Sergi},
        title = "{Modern stellar spectroscopy caveats}",
      journal = {\mnras},
         year = 2019,
        month = jun,
       volume = {486},
       number = {2},
        pages = {2075-2101},
          doi = {10.1093/mnras/stz549},
archivePrefix = {arXiv},
       eprint = {1902.09558},
 primaryClass = {astro-ph.SR},
       adsurl = {https://ui.adsabs.harvard.edu/abs/2019MNRAS.486.2075B}
}

@INPROCEEDINGS{Coule2003,
       author = {{Coud{\'e} du Foresto}, Vincent and {Borde}, Pascal J. and {Merand}, Antoine and {Baudouin}, Cyrille and {Remond}, Antonin and {Perrin}, Guy S. and {Ridgway}, Stephen T. and {ten Brummelaar}, Theo A. and {McAlister}, Harold A.},
        title = "{FLUOR fibered beam combiner at the CHARA array}",
    booktitle = {Interferometry for Optical Astronomy II},
         year = 2003,
       editor = {{Traub}, Wesley A.},
       series = {Society of Photo-Optical Instrumentation Engineers (SPIE) Conference Series},
       volume = {4838},
        month = feb,
        pages = {280-285},
          doi = {10.1117/12.459942},
       adsurl = {https://ui.adsabs.harvard.edu/abs/2003SPIE.4838..280C}
}

@ARTICLE{Wittkoski2004,
       author = {{Wittkowski}, M. and {Aufdenberg}, J.~P. and {Kervella}, P.},
        title = "{Tests of stellar model atmospheres by optical interferometry. VLTI/VINCI limb-darkening measurements of the M4 giant {\ensuremath{\psi}} Phe}",
      journal = {\aap},
         year = 2004,
        month = jan,
       volume = {413},
        pages = {711-723},
          doi = {10.1051/0004-6361:20034149},
archivePrefix = {arXiv},
       eprint = {astro-ph/0310128},
 primaryClass = {astro-ph},
       adsurl = {https://ui.adsabs.harvard.edu/abs/2004A&A...413..711W}
}

@ARTICLE{Wittkowski2006,
       author = {{Wittkowski}, M. and {Aufdenberg}, J.~P. and {Driebe}, T. and {Roccatagliata}, V. and {Szeifert}, T. and {Wolff}, B.},
        title = "{Tests of stellar model atmospheres by optical interferometry. IV. VINCI interferometry and UVES spectroscopy of Menkar}",
      journal = {\aap},
         year = 2006,
        month = dec,
       volume = {460},
       number = {3},
        pages = {855-864},
          doi = {10.1051/0004-6361:20066032},
archivePrefix = {arXiv},
       eprint = {astro-ph/0610150},
 primaryClass = {astro-ph},
       adsurl = {https://ui.adsabs.harvard.edu/abs/2006A&A...460..855W}
}

@ARTICLE{Aufdenberg2005,
       author = {{Aufdenberg}, J.~P. and {Ludwig}, H.-G. and {Kervella}, P.},
        title = "{On the Limb Darkening, Spectral Energy Distribution, and Temperature Structure of Procyon}",
      journal = {\apj},
         year = 2005,
        month = nov,
       volume = {633},
       number = {1},
        pages = {424-439},
          doi = {10.1086/452622},
archivePrefix = {arXiv},
       eprint = {astro-ph/0507336},
 primaryClass = {astro-ph},
       adsurl = {https://ui.adsabs.harvard.edu/abs/2005ApJ...633..424A}
}

@ARTICLE{Lindegren2021,
       author = {{Lindegren}, L. and {Bastian}, U. and {Biermann}, M. and {Bombrun}, A. and {de Torres}, A. and {Gerlach}, E. and {Geyer}, R. and {Hern{\'a}ndez}, J. and {Hilger}, T. and {Hobbs}, D. and {Klioner}, S.~A. and {Lammers}, U. and {McMillan}, P.~J. and {Ramos-Lerate}, M. and {Steidelm{\"u}ller}, H. and {Stephenson}, C.~A. and {van Leeuwen}, F.},
        title = "{Gaia Early Data Release 3. Parallax bias versus magnitude, colour, and position}",
      journal = {\aap},
         year = 2021,
        month = may,
       volume = {649},
          eid = {A4},
        pages = {A4},
          doi = {10.1051/0004-6361/202039653},
archivePrefix = {arXiv},
       eprint = {2012.01742},
 primaryClass = {astro-ph.IM},
       adsurl = {https://ui.adsabs.harvard.edu/abs/2021A&A...649A...4L}
}

\begin{appendix}

\section{Plots of fitting}

For HD\,222368 and HD\,19787, which were observed as part of the ISSP programme, we compare the effect of including the correlation between the calibrators and transfer functions as described in Sect.~\ref{sec:2.4}. These comparisons are summarised in Table~\ref{tab:table4}.

\begin{figure*}[ht!]
    \centering
    
    \includegraphics[width = 17 cm]{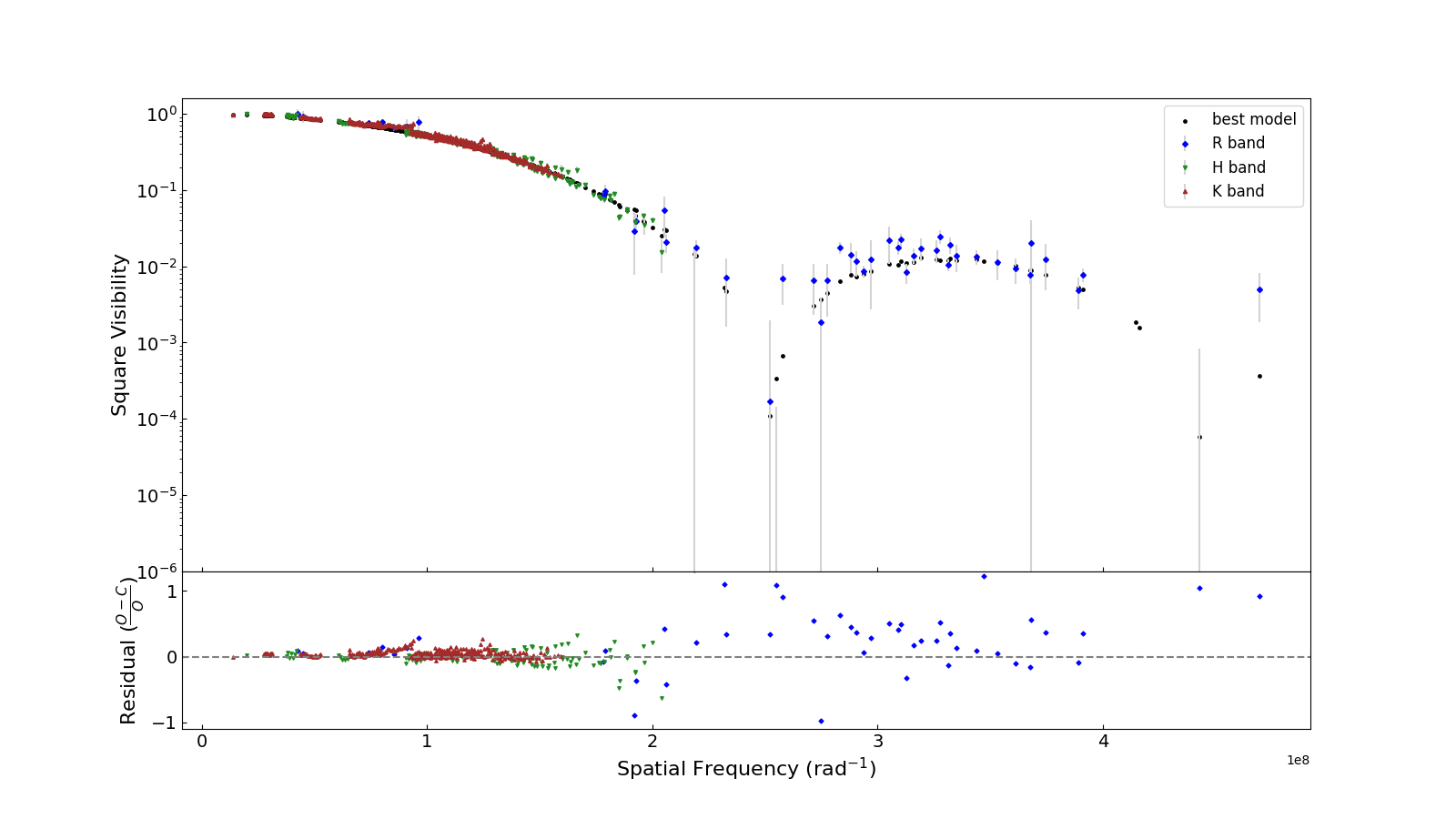}
    \vspace{-0.7cm}

    \includegraphics[width = 17 cm]{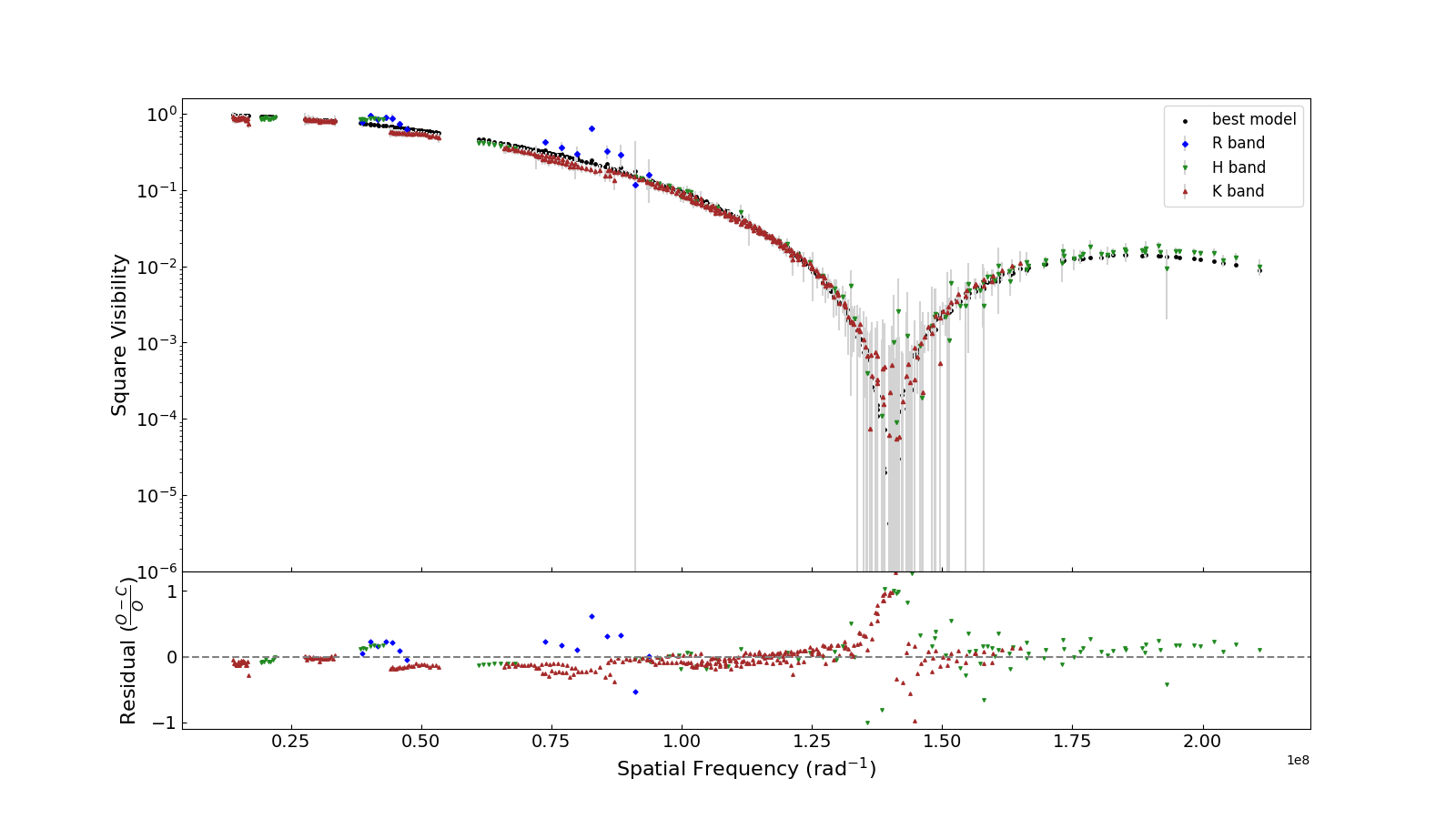}

    \caption{ Square visibility fits for $\iota$ Psc (HD~222368, top) and for $\delta$ Ari (HD~19787, bottom) without correlation using \texttt{PIMS}. The legends in the plot are the same as in Fig.~\ref{fig:fig2}.}
    \label{fig:figA1}
\end{figure*}

The square visibilities fitted with PIMS, including the correlation matrix in the fit, are shown in Figure~\ref{fig:fig2} and \ref{fig:fig3}, respectively. The figure~\ref{fig:figA1} shows the figure from the fit of the data without using the correlation matrix while fitting. The top panel shows the plot of HD\,222368, and the bottom panel shows that of HD\,19787. In these figures, the black points represent the modelled square visibilities from the ANN in the R, H, and K bands, while the dark blue, green, and maroon points correspond to the observations from SPICA (R band), MIRC-X (H band), and MYSTIC (K band), respectively. Comparing the fits and residuals in both cases, no significant visual differences are evident. However, differences can be seen in the reduced $\chi^2$ values and in the fitted stellar parameters.

\begin{figure*}[ht!]
    \centering
    
    \includegraphics[width = 17 cm]{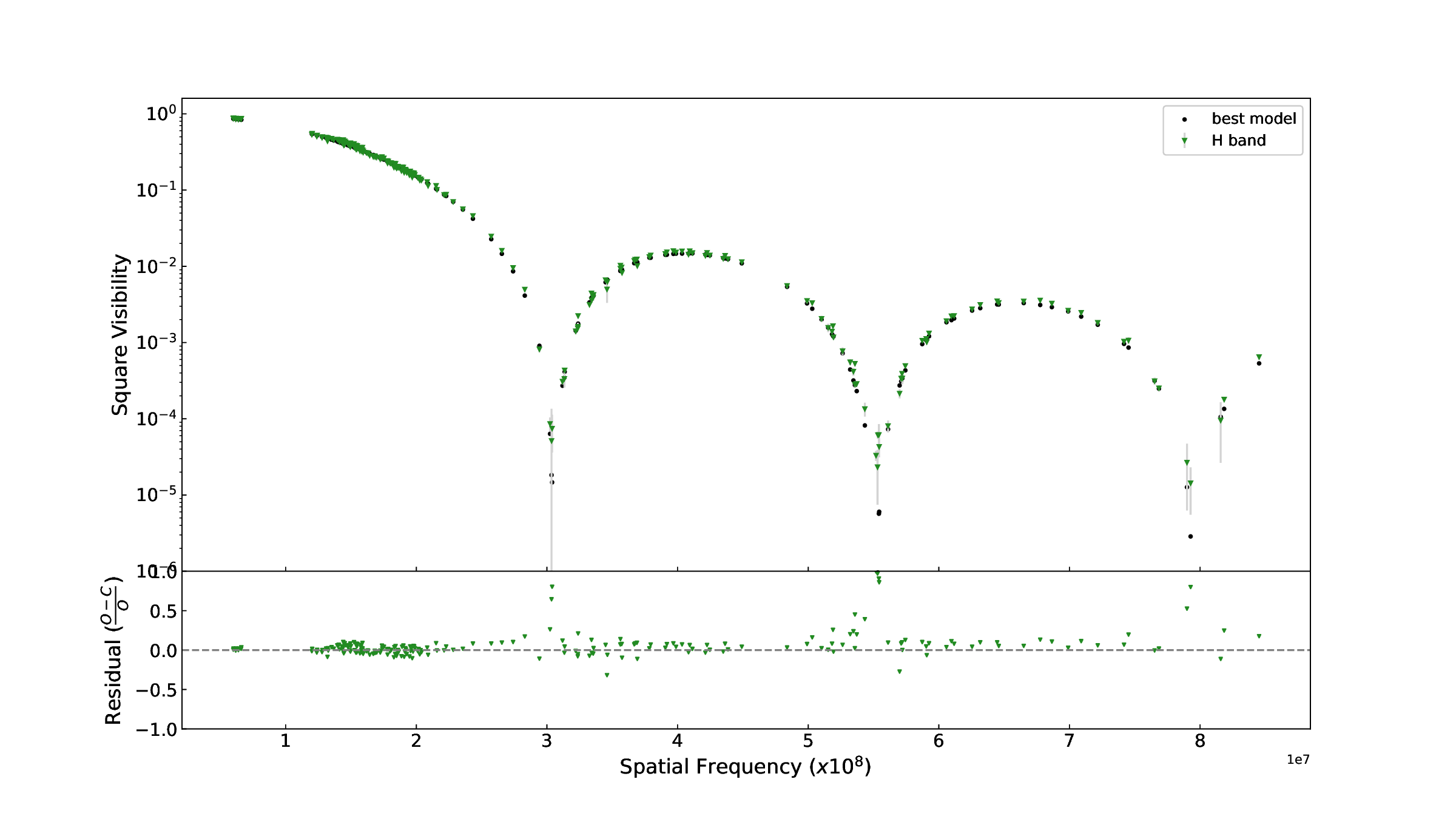}
    \vspace{-0.7cm}

    \includegraphics[width = 17 cm]{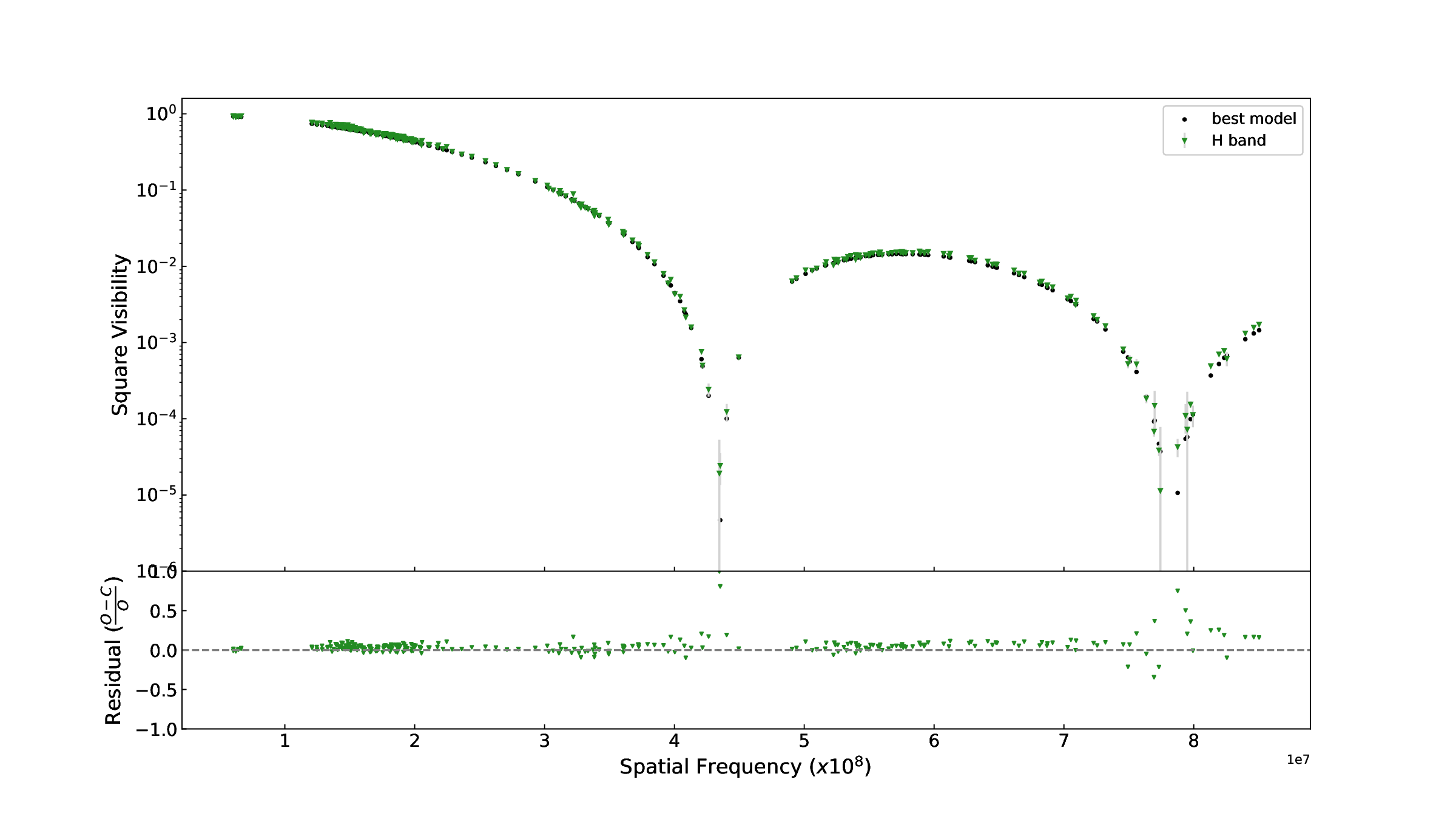}

    \caption{ Square visibility fits for $\alpha$ Cen A (HD 128620) (top) and $\alpha$ Cen B (HD 128621) (bottom) using \texttt{PIMS}. The dark green points represent the observations and the light green are the best fitted model. The grey lines indicate the observational uncertainties.}
    \label{fig:figA2}
\end{figure*}

Similarly, Figures~\ref{fig:figA2} and \ref{fig:figA4} show the fitted square visibilities and the corner plots for the $\alpha$ Cen targets observed with VLTI/PIONIER. Since insufficient information was available to estimate the correlation matrix for these targets, they could not be fitted with the correlation matrix.
 
\begin{figure*}[ht!]
    \centering

    \begin{subfigure}[t]{0.45\textwidth}
        \centering
        \includegraphics[width=\textwidth]{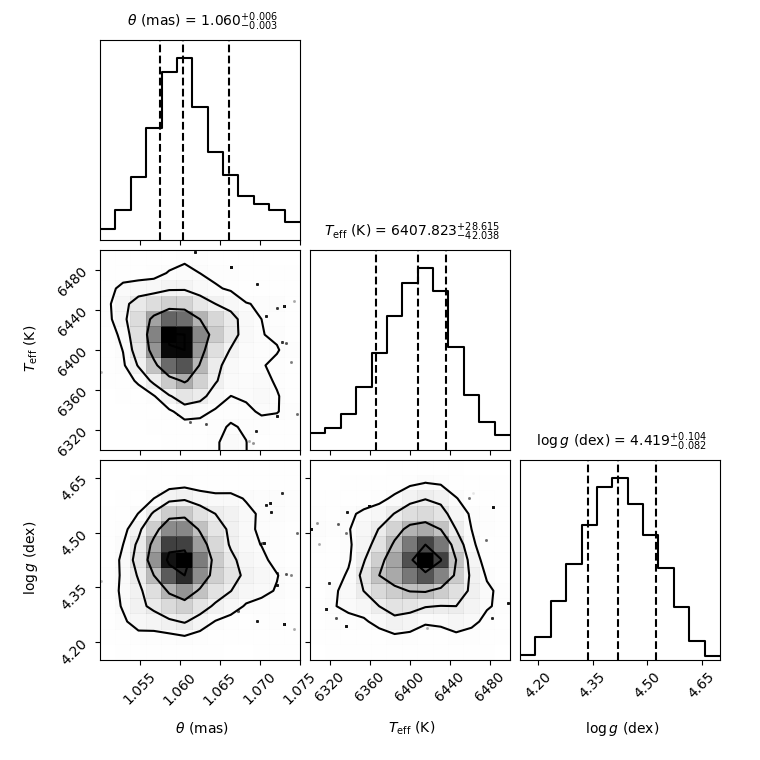}
        \caption{The corner plot for the $\iota$ Psc (HD222368) without the correlation between different baselines of the observations for both calibrators and targets throughout the night.}
    \end{subfigure}
    \hfill
    \begin{subfigure}[t]{0.45\textwidth}
        \centering
        \includegraphics[width=\textwidth]{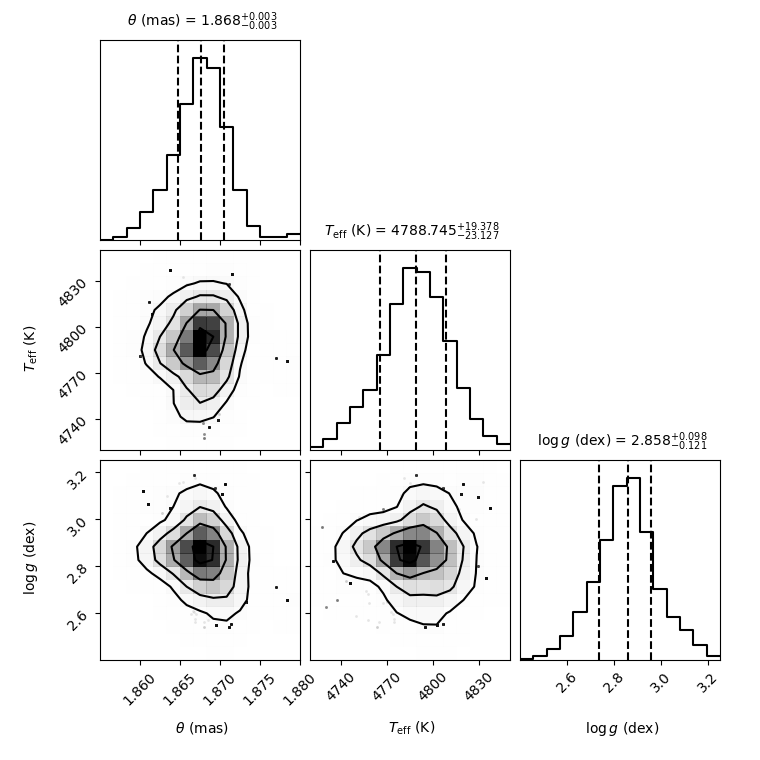}
        \caption{The corner plot for the $\delta$ Ari (HD~19787) without the correlation between different baselines of the observations for both calibrators and targets throughout the night.}
    \end{subfigure}

    \caption{Corner plots for HD 222368 and $\delta$ Ari (HD19787) from the ISSP observations without the correlation matrix using \texttt{PIMS}. }
    \label{fig:figA3}
\end{figure*}

\begin{figure*}[ht!]
    \centering

    \begin{subfigure}[t]{0.45\textwidth}
        \centering
        \includegraphics[width=\textwidth]{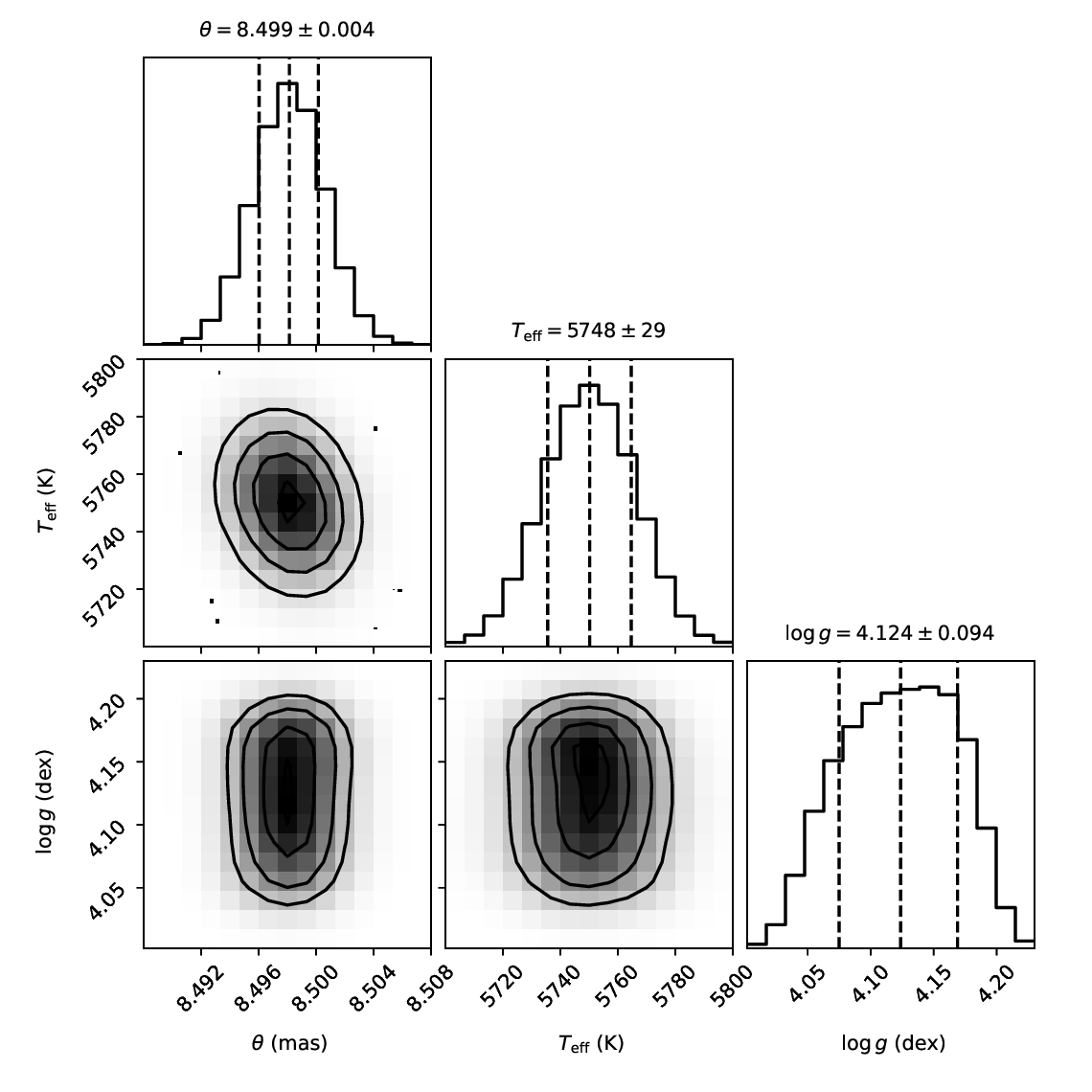}
        \caption{The corner plot for the $\alpha$ Cen A (HD 128620).}
    \end{subfigure}
    \hfill
    \begin{subfigure}[t]{0.45\textwidth}
        \centering
        \includegraphics[width=\textwidth]{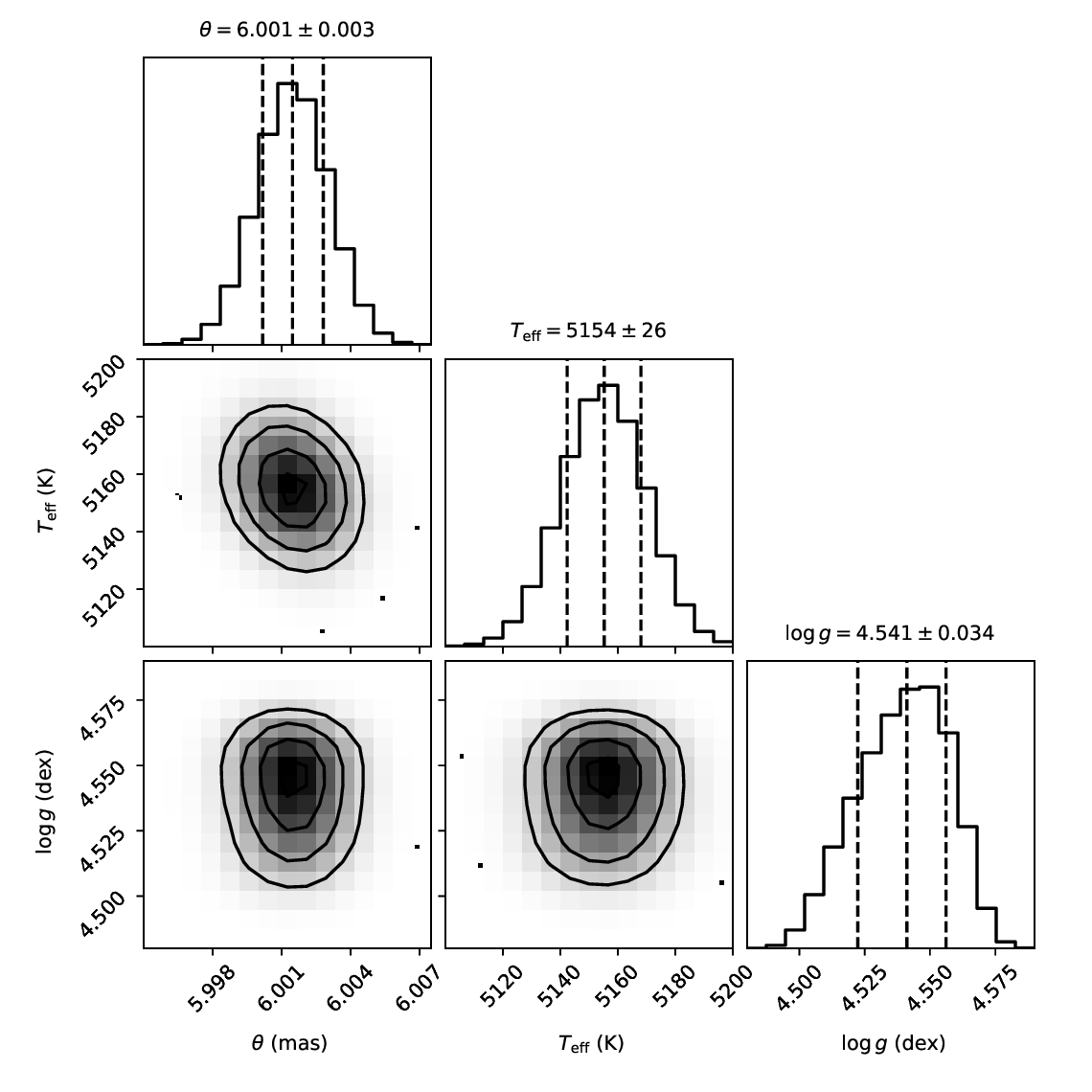}
        \caption{The corner plot for the $\alpha$ Cen B (HD 128621).}
    \end{subfigure}    \label{fig:figA5}
\caption{The corner plots of the $\alpha$ Cen stars from VLTI/PIONIER observations using \texttt{PIMS}.}
\label{fig:figA4}
\end{figure*}

The corner plot from the MCMC sampling for the ISSP target with a correlation matrix was depicted in Figure~\ref{fig:fig4}. Figure~\ref{fig:figA3} presents the corner plots obtained from the MCMC sampler for the fits without including the correlation matrix. The top-left panel shows the corner plot with the fit of the angular diameter, $\teff$, and $\logg$ for HD,222368, while the top-right panel shows the corresponding plot for HD,19787. Similarly, the corner plots for the $\alpha$ Cen targets are shown in Figure~\ref{fig:figA4}.

\label{sec:AppA}

\end{appendix}

\end{document}